\documentclass[11pt]{article}
\usepackage{epsf,cite,amssymb}

\newcommand{\resection}[1]{\setcounter{equation}{0}\section{#1}}
\newcommand{\al}{\alpha}

\newcommand{\be}{\begin{equation}}
\newcommand{\ee}{\end{equation}}
\newcommand{\eq}{\begin{equation}}
\newcommand{\en}{\end{equation}}
\newcommand{\ba}{\begin{array}}
\newcommand{\ea}{\end{array}}
\newcommand{\bea}{\begin{eqnarray}}
\newcommand{\eea}{\end{eqnarray}}

\newcommand{\bc}[6]{{\,C^{(\!#1\hspace{-0.3pt}#2\hspace{-0.3pt
     }#3\!)#6}_{\;#4#5}\,}}
\newcommand{\bch}[6]{\ensuremath{\,{\hat C}^{(\!#1\hspace{-0.3pt}#2\hspace{-0.3pt}#3\!)#6}_{\;#4#5}\,}}
\newcommand\bcs[4]{{C^{(\,#1\,)}_{#2#3#4}}}
\newcommand{\bp}[3]{\ensuremath{\,\f^{(\!#1\hspace{-0.7pt}#2\!)}_{#3}}}
\newcommand{\bph}[3]{\ensuremath{\,\widehat\varphi^{(\!#1\hspace{-0.7pt}#2\!)}_{#3}}}

\newcommand{\blank}[1]{}

\newcommand{\cev}[1]{\langle \,#1\,|}
\newcommand{\ceq}{{\ensuremath{{c\,{=}1}}}}
\newcommand{\clim}{{\ensuremath{c\to 1}}}
\newcommand{\clt}{{\ensuremath{{c\,{<}1}}}}
\def\cH{{\mathcal{H}}}
\newcommand{\nH}[1]{{\ensuremath{{\cal H}^{(#1)}}}}
\newcommand{\cN}{{\ensuremath{\mathcal{N}}}}

\newcommand{\D}{{{\rm d}}}
\newcommand{\ds}{\displaystyle}

\newcommand{\elim}{{\ensuremath{\eps\to 0}}}
\newcommand{\eps}{\epsilon}
\newcommand{\eref}[1]{(\ref{#1})}

\newcommand{\f}{{\phi}}
\newcommand{\ftt}{{\ensuremath{\f_{33}}}}

\newcommand{\fract}[2]{{\textstyle\frac{#1}{#2}}}
\newcommand{\F}[6]{\ensuremath{%
\textsf{\large F}^{}_{#5#6}\!\left[
     \begin{array}{ll}
     \!#2 & \!#3\! \\ \!#1 & \!#4\! 
     \end{array}\!\right]}}
\newcommand{\Fh}[6]{\ensuremath{\mbox{\ensuremath{%
\widehat{\textsf{\large  F}}^{}_{#5#6}\!\left[
     \begin{array}{ll}
     \!#2 & \!#3\! \\ \!#1 & \!#4\! 
     \end{array}\!\right]}}}}

\renewcommand{\hat}{\widehat}
\newcommand{\hP}{{\hat P}}
\renewcommand{\hP}{{P}}

\newcommand{\Id}{\ensuremath{{1\hspace{-4.5pt}1}}}

\newcommand{\nn}{\nonumber}

\newcommand{\One}{{\hbox{{\rm 1{\hbox to 1.5pt{\hss\rm1}}}}}}
\renewcommand{\One}{{\mathbb 1}}
\renewcommand{\One}{{\rm 1\!\!1}}
\def\OPE{ope}
\def\OPES{opes}

\newcommand{\sm}[2]{S_{#1}{}^{#2}}
\newcommand{\Tr}{{\mathrm Tr}}
\newcommand{\ts}{\textstyle}

\newcommand{\vac}{\vec 0}
\renewcommand{\vec}[1]{|\,#1\,\rangle}
\newcommand{\vev}[1]{\langle\,#1\,\rangle}

\newcommand{\bfb}[3]{
     \raisebox{-1.45mm}{
     \setlength{\unitlength}{0.85mm}     
     \begin{picture}(14,5)(5,0)
          \put(3,1){\line(1,0){6.3}}       
          \put(10.7,1){\line(1,0){6.3}}
          \put(10,1){\qbezier(-0.7,0)(-0.7,0.7)(0,0.7)
               \qbezier(0,0.7)(0.7,0.7)(0.7,0)}          
          \put(6.5,2){\makebox(0,0)[cb]{\small\ensuremath{\smash{#1}}}}
          \put(13.5,2){\makebox(0,0)[cb]{
               \small\ensuremath{\smash{#3}}}}
          \put(10,3.5){\makebox(0,0)[cb]{
               \small\ensuremath{\smash{#2}}}}
     \end{picture}}}

\newcommand{\tfrac}{\fract}

\begin{document}

\begin{titlepage}
\vskip 0.5cm
\begin{flushright}
KCL-MTH-01-01 \\
PAR-LPTHE-01-01\\
{\tt hep-th/0101187}\\
\end{flushright}
\vskip 4.cm
\begin{center}
{\Large {\bf 
Minimal model boundary flows and \ceq\ CFT
}} \\[5pt]
\end{center}
\vskip 1.3cm
\centerline{K.~Graham%
\footnote{e-mail: {\tt kgraham@mth.kcl.ac.uk}},
I.~Runkel\footnote{e-mail: {\tt ingo@lpthe.jussieu.fr}}
and G.M.T.~Watts\footnote{e-mail: {\tt gmtw@mth.kcl.ac.uk}}
}
\vskip 0.6cm
\centerline{${}^{1,3}$\sl Mathematics Department, }
\centerline{\sl King's College London, Strand, London WC2R 2LS, U.K.}
\vskip 0.4cm
\centerline{${}^2$\sl LPTHE, Universit\'e Paris VI,}
\centerline{\sl 4 place Jussieu, F\,--\,75\,252\, Paris\, Cedex 05, France}
\vskip 0.9cm
\begin{abstract}
\vskip0.15cm
\noindent
We consider perturbations of unitary minimal models by boundary
fields.
Initially we consider the models in the limit as \clim\ and find
that the relevant boundary fields all have simple interpretations in
this limit.
This interpretation allows us to conjecture the IR limits of flows in
the unitary minimal models generated by the fields $\f_{rr}$ of `low'
weight. 
We check this conjecture using the truncated conformal space approach.
In the process we find evidence for a new series of integrable
boundary flows. 

\end{abstract}
\end{titlepage}
\setcounter{footnote}{0}
\def\thefootnote{\fnsymbol{footnote}}

\resection{Introduction}

We consider perturbations of unitary minimal models by boundary fields.
Perturbations of these models by the least relevant boundary field
have been studied by Recknagel et al. 
\cite{RRSch1}, in perturbation theory around \ceq.
One simple observation is that all the perturbative flows they find
must collapse to identities at \ceq, 
(i.e.\ the start and end points of the flows for \clt\ must
be identified at \ceq)
and so one might hope to
gain some insight into flows for \clt\  generally by examining the
structure of the \clim\ limit of the boundary minimal models.
In particular, non-perturbative flows are hard to understand in
absence of a Landau Ginzburg picture, and they may be easier to
understand as `perturbations' of genuine \ceq\ flows.  
This paper provides arguments to support this point of view,
which was outlined in \cite{GRW1}.

In section \ref{sec:bumm} we consider the unperturbed boundary
conditions (b.c.'s).
After some generalities,  
we review the `A' type unitary minimal models and their
b.c.'s which we shall denote $(r,s)\equiv B_{(r,s)}$.
Of these, $(1,r)$ and $(r,1)$ play a special
role in the \clim\ limit, and the two b.c.'s become identified as a
new boundary condition which we denote by $(\hat r)$. We derive
various properties of the $(\hat r)$ b.c.'s. 
We can then formulate our main conjecture, that 
in the limit \clim, a general boundary condition of type $(r,s)$
splits up into a superposition of min~$(r,s)$ such `fundamental'
b.c.'s:
\be
 \lim_\clim \; 
 (r,s)
=
  \oplus_{n=1}^{\min(r,s)}
  \;
  (\, \hat{|r{-}s|{+}2n{-}1} \,)
\;.
\label{eq:conj0}
\ee

Superpositions of boundary conditions have already been seen in the
form of Chan-Paton factors in string theory (see e.g. \cite{Pol}) and
also in the work of Affleck \cite{Affleck} and in \cite{RRSch1}.
The superpositions generically contain multiple boundary fields of
weight zero. In our case we can investigate the properties of these
weight zero
fields explicitly since they are the \clim\ limit of 
the primary fields $\f_{rr}$ in the unitary minimal models, 
and we know all their structure constants from \cite{Runk1}.

The weight zero and weight one fields on a boundary with b.c.\
(\ref{eq:conj0}) are of special interest since these include the
limits of all the 
relevant boundary fields for \clt.
We argue that the weight zero fields on the \clim\ limit of the
$(r,s)$ b.c.\ can be decomposed into linear  combinations of
projectors onto the fundamental boundary conditions 
$(\hat t)$ appearing in (\ref{eq:conj0}).
The weight one fields can then be divided into fields which act
solely within the fundamental sectors, and fields which interpolate
pairs of them  (i.e.\ boundary condition changing operators). 
We present an explicit analysis of the $(2,2)$ boundary condition in
section \ref{sec:22}, and for the series of boundary conditions
$(2,p)$ and $(3,p)$ in appendix \ref{app}.

In section \ref{sec:c<1} we turn to boundary perturbations, and start
with the case of \ceq. 
Since  the weight zero fields can be expressed in terms of projectors,
perturbation by these fields becomes easy to understand.
We then turn to the models with \clt.
In section \ref{sec:c<1b} we examine the perturbations by the fields
of type $\f_{rr}$ using the
truncated conformal space approach (TCSA) and find that we are indeed
able to predict the IR end-points from our analysis of the
corresponding flows at \ceq\  
(up to the ambiguity $(r,s)\leftrightarrow(s,r)$ in the boundary conditions).
In the process, we find strong evidence for the integrability of the
perturbation of the model $M_{r+1,r+2}$ by the particular 
boundary
field $\f_{rr}\equiv\f_{12}$. 

The perturbations of the \ceq\ model
by the weight one fields is more difficult and we defer this to a
later paper \cite{next}, along with a discussion of perturbations of
the minimal models by boundary fields of type $\f_{r,r+2}$. We
conclude with some discussion of these results and possible
extensions. 

\resection{The boundary conditions of the unitary minimal models}
\label{sec:bumm}

The original papers by Cardy and Lewellen 
\cite{Card4,CLew1,Lewe1} on boundary conformal field
theory set out the basic properties --
the boundary field content, and the consistency conditions satisfied by
the various structure constants.
The boundary field contents of all Virasoro minimal
models were found 
in \cite{BPPZ1,BPPZ2,BPPZ3}, and  a full solution of these consistency
conditions for the $A$--type 
Virasoro minimal models was proposed in \cite{Runk1}.
For more recent developments, see \cite{PZube,CS}.
First we give some general results on boundary models and then discuss
the minimal models for \clt.

\subsection{Some generalities}

Consider the upper half plane (UHP) with the boundary
condition $\alpha$ on the left real axis and $\beta$ on the right real
axis.  A single copy of the Virasoro algebra acts on the upper half
plane, and so the Hilbert space $\cH_{\alpha\beta}$ of the upper half
plane with this pair of boundary conditions splits into a direct sum
of irreducible representations $R_c$ of the Virasoro algebra%
\footnote{In general this can only be proven for unitary models},
\be
  \cH_{\alpha\beta}
= {\textstyle \sum_c}\; n_{\alpha\beta}{}^c\; R_c
\;,
\label{(a)}
\ee
where the numbers $n_{\alpha\beta}{}^c$ should be non-negative integers;
if we further impose the condition that the identity
representation $R_{\One}$ appear at most once, 
we call the resulting subset of boundary conditions `fundamental'.

The states in the Hilbert $\cH_{\alpha\beta}$ are in one-to-one
correspondence with the fields which interpolate the boundary
conditions $\alpha$ and $\beta$. 
In particular, the fields which can lie on the $\alpha$ boundary are
in one-to-one correspondence with the states in $\cH_{\alpha\alpha}$,
and the space of primary fields of type $c$ which live on the boundary
$\alpha$ has dimension $n_{\alpha\alpha}{}^c$. 

The UHP can be related to an infinite strip of width $R$ by a
conformal transformation, and the Hamiltonian 
generating translations along the strip is 
(in terms of the Virasoro algebra on the UHP)
\be
  H(R) = (\pi/R)(L_0 - c/24)
\;.
\ee
Hence the partition function on a cylinder of width $R$ and
circumference $L$ with b.c.'s $(\alpha,\beta)$ on the two
edges is  
\be
  Z_{\alpha\beta}(L,R)
\equiv 
  {\rm Tr}_{\cH_{\alpha\beta}}\Big(\; e^{-LH(R)}\;\Big)
= {\textstyle \sum_c} \; n_{\alpha\beta}{}^c \; \chi^{}_c(q)
\;,
\label{(b)}
\ee
where $\chi^{}_c(q)$ are the characters of the irreducible Virasoro
highest weight representations $R_c$
\be
  \chi^{}_c(q)
= \Tr_{R_c} \Big(\; q^{L_0 - c/24} \Big)
\;,\;\;\;\;
  q 
= \exp( -\pi L/R )
\;.
\ee
Hence from equations \eref{(a)} and \eref{(b)}
we see that the cylinder partition functions encode the boundary field
content. 

\subsection{Minimal models}

From now on we shall assume that we are dealing with the $A$--type
Virasoro minimal models
(for general properties of the minimal models, see e.g. \cite{Ybk}).
Each model $M(p,p')$ is labelled by two positive coprime integers
$p,p'>1$, or alternatively by the rational number $t = p/p'$.
Associated to each model is a set of $(p-1)(p'-1)/2$ Virasoro
representations $R_i$ labelled by $i$, given by a set of weights $\{h_i\}$. 
The central charge $c$ and the weights $h_{rr'}$ take the form
\be
  c
= 13 - 6 t - 6/t
\;,\;\;\;\;
  h_{r{r'}}
= \frac{1}{4t}(\, (r {-}{r'}t)^2 \,-\, (1{-}t)^2\, )
\;,\;\;\;\;
\ee
where the Kac-labels $(r,{r'})$ lie in the ranges $r=1..p{-}1$ and
${r'}=1..p'{-}1$. 
If we allow all pairs $(r,{r'})$ in these ranges then each weight
appears twice, since $h_{r{r'}}=h_{p-r,p'-{r'}}$.

We can take a definite choice of representatives of the Kac labels
$(r,r')$ as follows.
At least one of $p$ and $p'$ is odd. Suppose $p$ is odd.
Then the pairs 
$\{$ $(r,r')$, $1\leq r \leq p-2$, $1\leq r' \leq p'-1$, $r$ odd $\}$
run over the set of Virasoro representations once and once only.

The fusion product of the representations $R_a$ is described by the
Verlinde algebra 
\be
   R_a \; \times \; R_b \; 
= \; {\textstyle \sum_c}\; N_{ab}{}^c \; R_c
\;,
\ee
where (in this case) the Verlinde fusion numbers 
$N_{ab}{}^c$ are either 0 or 1.
With the choice of representatives given above, the fusion numbers are
explicitly 
\be
 \begin{array}{l}
   N_{(rr')(ss')}{}^{(tt')}
 = \cN_{rs}^{t}(p)   \cdot \cN_{r's'}^{t'}(p') 
\\[2mm]
   \cN_{ab}^c(m) 
 = \cases{
   1 :
&  
   $ |a{-}b|<c<\min(a{+}b,\,2m{-}a{-}b) \;,\;
      a{+}b{+}c \hbox{ odd } $ \cr
   0 :
&  {otherwise} }
\end{array}
\label{eq:rule0}
\ee
The characters of the minimal model representations are
\bea
  \chi^{}_{(r,r')}(q)
&=& \frac {q^{-c/24}}{\varphi(q)}\,
  { \sum_{n=-\infty}^\infty}
\left(
  q^{ h_{ (r+2n p), r'} }
- q^{ h_{ (r+2n p),-r'} }
\right)
\;,\;\;
  \varphi(q)
= { \prod_{n=1}^\infty}(1 - q^n)
\;.
\label{eq:mmchars}
\eea
Their behaviour under modular transformations 
$q{=}e^{2\pi i\tau} \rightarrow \tilde q{=}e^{-2\pi i/\tau}$
is given by the matrix
\be
  S_{rr'}{}^{ss'} 
= 2^{3/2} (pp')^{-1/2} (-1)^{1+rs'+r'\!s}
  \sin(\pi rs/t) \sin(\pi r'\!s' t) 
\;.
\ee

\subsection{Boundary conditions}

For the A--type minimal models, the fundamental boundary conditions
are in 1--1 correspondence with the representations of the Virasoro
algebra and so we can label both boundary conditions and
representations from the same set $\{a\}$; in this case
the numbers $n_{ab}{}^c$ are the Verlinde fusion numbers
\cite{Card4}. 
Since the fundamental boundary conditions $a$ are in one-to-one
correspondence with the set of Virasoro representations, we will
denote them by $a$, $h_a$ or $(r^{}_a, r'_a)$ interchangeably.

It is convenient to define an ordering on the boundary conditions,
for example
\be
 (r,r')>(s,s') 
 \;\; \Leftrightarrow \;\;
 (r'{>}\,s' 
    \;\hbox{ or }\; 
 (r'{=}\,s' \hbox{ and }
    r\,{>}\,s))
\;.
\ee 
We choose to normalise the primary fields and one-point functions of
the boundary theory so that for boundary conditions $a,b$ and a
primary boundary field $i$ we have
\be
  \vev 1^a = \sm 1a/\sm 11 \;,\quad a\le b\,:\; \bc abaii1 = 1
\label{mm:norm}
\ee
In particular this implies that for
$a>b\,,\; \bc abaii1 = \sm 1b/\sm 1a$ \cite{Lewe1,Runk1}.

\resection{The \clim\ limit of the boundary unitary minimal models}
\label{sec:c=1}

To study the limit $c\rightarrow 1$, we set $t=1-\eps$, so that the 
central charge $c\sim 1-6\eps^2$. 
We shall denote characters, partition functions, Hilbert spaces,
etc, at \ceq\ by $\hat\chi$, $\hat Z$, $\hat \cH$, etc, to distinguish
them from those in the minimal models with \clt.
We first recall the result of Recknagel et al.\ \cite{RRSch1}
which led to our conjecture.

In \cite{RRSch1}, they studied the renormalisation group
flows of a boundary condition $(\al)\,{=}\,(a,a')$ generated by the 
field $\f_{13}^{(\al\al)}$ by the addition to the action of the
integral along the boundary,
\be
  \delta S
= \lambda\, 
  \oint 
  {\D l}\;\; 
  \f_{13}^{(\al\al)}
\;.
\label{eq:S'}
\ee
They calculated the perturbative beta function for $\lambda$ for
$\eps\sim 0$ and found an attractive fixed point for a value
$\lambda^* = O(\eps)$. 
They examined the properties of this fixed point and 
found that it could only be described by a superposition of b.c.'s,
\be
  (a,a')
\;
\mathop{\longrightarrow}\limits_{\lambda \to \lambda^*}
\;
  \oplus_{n=1}^{\min(a,a')}
  \;
  (\, |a{-}a'|{+}2n{-}1, 1 \,)
\;.
\label{eq:RRS}
\ee
Since the fixed point is at $\lambda^*=O(\eps)$, as \clim\ (and
consequently \elim) the fixed
point moves closer to $\lambda=0$, and in the limit this flow must
degenerate to an identity. 
Denoting the limit as $\clim$ of the 
b.c.\ $(r,1)$ by $(\hat r)$, 
we are led to our main conjecture:

{\narrower\parindent 0pt\parskip 4pt

{\bf Conjecture 1}

The \clim\ limit of the boundary condition $\al = (a,a')$ is the
superposition of $\min(a,a')$  `fundamental' boundary conditions,
which we represent schematically as
\be
 \lim_\clim \; 
 (a,a')
=
  \oplus_{n=1}^{\min(a,a')}
  \;
  (\, \hat{|a{-}a'|{+}2n{-}1} \,)
\;.
\label{eq:conj}
\ee
That is, the field content and the correlation functions for the
$(a,a')$ boundary condition are identical (in the \clim\ limit) to
those on the superposition of the fundamental boundary conditions.

}

\vskip 4pt
\noindent 
An important consequence of this conjecture is that the boundary
fields that arise on the $(r,s)$ b.c.\ in the \clim\ limit can be
expressed in terms of the boundary fields of the 
superposition. In particular, the scalar (weight zero) fields
that arise must be spanned by the projectors onto the various
fundamental components of the superposition (\ref{eq:conj}),
and the weight one fields must split into boundary fields living on a
single fundamental component and b.c.-changing fields which
interpolate two different fundamental b.c.'s.
This leads to:

{\narrower\parindent 0pt\parskip 4pt

{\bf Conjecture 2}

The operator product algebra $B$ of the 
scalar fields on the boundary condition 
$\al = \oplus \hat\al_i$ is 
the algebra of projectors $P_i$ onto the fundamental components 
$\hat\al_i$ of the
boundary condition $\al$.
In particular, if we consider the upper half plane, the scalar fields
on the left and right of the origin generate two commuting copies of this
algebra, $B^L$ and $B^R$, which act on the Hilbert space
$\hat\cH(\bfb \al{}\al)$, and  the projectors in these two algebras
project onto the subspaces   
\be
  P^L_i P^R_j 
: \hat\cH(\bfb \al{}\al) \longrightarrow
  \hat\cH(\bfb{ \hat \al_i}{}{ \hat \al_j}) \;.
  \label{H-decomp}
\ee
\par
}

\vskip 4pt
\noindent%
We have not been able to prove these conjectures, but have checked them
quite extensively.
In the next section we show that the
\clim\ limits of the cylinder partition functions
are in agreement with conjecture 1, and 
in section \ref{sec:22} we find explicitly the relation between the
weight zero and weight one boundary fields on the b.c.\ $(2,2)$ and
the superposition $(\hat 1)\oplus(\hat 3)$, and show that they are in
accord with conjecture 2.
We have also examined the general $(2,p)$ and $(3,p)$ boundaries, and
leave these results to the appendix in sections 
\ref{app:2p} and \ref{app:3p}.

\subsection{The \clim\ limit of the cylinder partition functions}

The limit \elim\ corresponds to picking a
sequence of minimal models such that the ratio $p/p'$ approaches $1$. 
For the bulk models this poses some problems, since the bulk theory
has $(p-1)(p'-1)/2$ fields, which clearly tends to infinity as
\clim.
However, the boundary theories are rather better behaved in this
respect.

One important feature of the limit \clim\ which we note here is that
for $p$ and $p'$ large enough, the truncated fusion rules
(\ref{eq:rule0}) are replaced by simple $su(2)$ fusion rules:
\be
 \hat\cN_{ab}{}^c
\equiv
 \lim_{m\to\infty}
 \cN_{ab}{}^c(m)
= \cases{
   1 :
&  
   $ |a{-}b|<c<(a{+}b) \;,\;
      a{+}b{+}c \hbox{ odd } $ \cr
   0 :
&  {otherwise} }
\label{eq:rule1}
\ee

Let us now consider one or more particular fixed boundary conditions
\hbox{$\al{=}(a,a')$}, 
\hbox{$\beta{=}(b,b')$}, etc. 
For fixed boundary conditions $\alpha,\beta$ and 
$p$ and $p'$ large enough, the untruncated fusion rules (\ref{eq:rule1})
mean we can write the partition function $Z_{(\alpha,\beta)}$ as
\be
  Z_{(a,a'),(b,b')}
= \sum_{ c\in a\otimes b \atop c' \in a'\otimes b'}
  \chi^{}_{(c,c')}
\;,
\label{eq:Zid1}
\ee
where $c\in a \otimes b$ is a shorthand notation to indicate that the
sum runs over all labels $c$ that occur in the tensor product of the
$su(2)$ representations $a$ and $b$.
Hence, for $p$ and $p'$ large enough, 
there are $\min(a,b)\cdot\min(a',b')$ primary fields
interpolating the boundary conditions $\al$ and $\beta$,
and in particular the boundary condition
$\al$ has a fixed boundary field content of $a \cdot a'$ fields.
Furthermore, for $p,p'$ large enough, the number of states (up to any
particular level) becomes constant, and we can hope that any
particular physical quantity in the theory (structure constant,
correlator, etc) will approach a limiting value as \clim. 
We cannot prove that this limit is well defined, but
examination of several cases suggests that this is likely to be the case.
As a first step to finding this limit, we can find the
field content from the strip partition functions.

While the partition function itself has a smooth limit as \clim, the
decomposition of the Hilbert space into irreducible representations of
the Virasoro algebra does not.
The essential point is that the \elim\ limit of the minimal model
character $\chi^{}_{rr'}$ with fixed $r,r'$ is not, in general, the
character of an irreducible \ceq\ representation.
We shall discuss in section \ref{sec:irreps} how it is that 
the fields and states in a single irreducible representation for \clt\
can reassemble themselves into several irreducible representations at
\ceq. For the moment we shall assume this works, and present the results.

The relevant irreducible highest-weight representations at \ceq\ are
labelled by a single positive integer ($r$) with weights and characters 
\be
  \hat h_r 
= \frac{(r-1)^2} 4
\;,\;\;\;\;
  \hat \chi^{}_r
= \frac{q^{h_r-1/24}}{\varphi(q)}\,
  (\, 1 - q^{r} \,)
\;.
\label{eq:c=1}
\ee
In terms of these, we have
\be
  \lim_\clim\; h_{r{r'}} 
= \hat h_{|r-{r'}|+1}
\;,\;\;\;\;\;\;\;\;
  \lim_\clim \;  
  \chi^{}_{(r,{r'})}
= \sum_{n=1}^{\min(r,{r'})}  \;
  \hat\chi^{}_{|r-{r'}|+2n-1}
\;,
\label{eq:c=1b}
\ee
and in the limit \clim\ the representations $R_{(r,r')}$ and
$R_{(r',r)}$ are identical.
In particular,
\be
  \lim_\clim\; h_{1,r}^{}
= \lim_\clim\; h_{r,1}^{}
= \hat h_r^{}
\;,\;\;\;\;
  \lim_\clim\; \chi^{}_{1,r}
= \lim_\clim\; \chi^{}_{r,1}
= \hat \chi^{}_r
\;,
\label{eq:c=1c}
\ee
and so the representations $R_{(r,1)}$ and $R_{(1,r)}$ both have as
their limit the single irreducible representation $R_{(\hat r)}$,
justifying our notation $\lim_\elim (r,1) = (\hat r)$.

Note that the decomposition (\ref{eq:c=1b}) is given by the same
$su(2)$ fusion rules that appear in (\ref{eq:rule1}), so that we can
just as well write
\be
  \lim_\clim \;  
  \chi^{}_{(r,r')}
= \sum_{s\in r \otimes r'} \hat\chi^{}_{s}
\;.
\label{eq:Zid2}
\ee
Applying this to (\ref{eq:Zid1}),
we find
\be
  \lim_\clim \;  
  Z_{(a,a'),(b,b')}
= \!\!\sum_{ c\in a\otimes b \atop c' \in a'\otimes b'}
  \lim_\clim \;  
  \chi^{}_{(c,c')}
\,
= \!\!\sum_{ {d\in a\otimes b\otimes a'\otimes b'} \atop {\vphantom{a}}}
  \!\!\!\!\!\hat\chi^{}_{d}
\,
= \!\!\sum_{ {e\in a\otimes a' }\atop {e' \in b\otimes b'}}
  \sum_{ {d\in e \otimes e'} \atop{\vphantom{a}}}
  \!\hat \chi^{}_{d}
\,
= \!\!\sum_{ {e\in a\otimes a' }\atop {e' \in b\otimes b'}}
  \!\!\hat Z_{(\hat e)(\hat e')}
\;\,,
\label{eq:Zid3}
\ee
where  in the last equality we used equation (\ref{(b)}) as applied to
the \ceq\ fusion rules (\ref{eq:rule1}).
In other words, if we write the characters appearing in the
decomposition 
as
\be
  \lim_\clim \chi^{}_a
= \sum_{\{a_i\}}\, \hat \chi_{a_i}^{}
\;,
\label{eq:clim1}
\ee
then the partition function for the cylinder with boundary conditions
$(\al,\beta)$ satisfies
\be
   \lim_\clim Z_{\al,\beta}
= {\sum_{\{\al_i\},\{\beta_j\}}} \, 
  \hat Z_{ \al_i,  \beta_j}
\;,
\label{eq:z}
\ee
which is in exact agreement with the conjecture (\ref{eq:conj}).

\subsubsection{The `missing' fields}
\label{sec:irreps}

As explicitly shown in (\ref{eq:Zid2}), the limit of an irreducible
representation of the Virasoro algebra for \clt\  need not be an
irreducible representation at \ceq. This is due to to the fact that
certain vectors may become null as \clim.
Consider the representation $h_{rr'}$. For \clt\  and $p,p'$ large
enough, the first null vector in the representation occurs at level
$r{\cdot}r'$. However, at \ceq, $h_{rr'} = \hat h_{|r-r'|+1}$, and the
first null vector in this representation occurs at level
$l = (|r-r'|+1)$. 
Since we require the number of fields of a given weight not to
change abruptly as \clim, we need to find a way to keep this state in
the spectrum. The solution is simply to normalise the state to unit
norm so that it does not decouple, and normalise the corresponding
field accordingly. 
This may lead to
divergent correlation functions involving this new field, but we find
in practice that this is not the case.
For example, consider the case of the $(33)$ representation.
We have
\be
  \lim_{\clim}\, h_{33} 
\;=\;
  0
\;,\;\;\;\;
  \lim_{\clim}\,
  \chi^{}_{33}
\;=\;
 \hat \chi^{}_1 
\;+\;
 \hat \chi^{}_3
\;+\;
 \hat \chi^{}_5 
\;.
\ee
We see that at \ceq\ there arise two new primary fields
of weight $\hat h_3=1$ and $\hat h_5=4$,
and correspondingly there are also null states at level 1 in the
representation with $h=0$,  at level 3 in the representation with
$h=1$, and at level 5 in the representation with $h=4$.
Concentrating on the representation with $h=0$, the new null state at
level $1$ is simply 
\be
  L_{-1} \vec{ h_{33} }
\;.
\ee
If we now consider the same state for \clt, we find that its norm is
\be
  \cev{h_{33}} L_1 L_{-1} \vec{h_{33}}
= 2 h_{33}
= 4 \eps^2 / t
\;,
\ee
and it will decouple from all correlation functions.
However, if we define the state
\be
  \vec{d_3}
= \lim_{\elim}\,
  \frac{\sqrt{1{-}\eps}}{2\,\eps}\; 
  L_{-1}\vec{h_{33}}
\;,
\ee
this new state has unit norm, $\langle d_3 | d_3 \rangle = 1$, 
and has all the properties we require; in particular it is a highest
weight state, since it is annihilated by $L_m$ with $m>0$.
For example, the action of $L_1$ is
\be
  L_1 \left[\; \frac{\sqrt{1{-}\eps}}{2\epsilon}\;
  L_{-1} \vec { h_{33}} \; \right]
= \frac{\sqrt{1{-}\eps}}{2\epsilon} \, 
  (2 h_{33})\,\vec {h_{33}}
= 2\epsilon\,\vec{h_{33}} + O(\epsilon^2)
\;,
\ee
so that $L_1 \vec{d_3} = 0$.
We expect all the extra required primary fields to arise in this way,
and their correlation functions to be well defined in the \clim\
limit. For example, the \OPES\ of $d_3$ with the fields of weight 0
and 1 on the $(2,2)$ b.c.\ are calculated in the appendix and shown to
be regular.

\subsection{The scalar fields of weight $0$}
\label{sec:scalars}

For this section we will restrict ourselves to the
case of the upper half plane with a single boundary condition
$\al=(a,a')$.
For general $c$, the Hilbert space 
$\cH_{\al\al}$ splits as
\[
  \cH_{\al\al} = \ts \sum_k N_{\al\al}{}^\kappa\, R_\kappa
\;,
\]
so that only representations $\kappa$ with $N_{\al\al}{}^\kappa$
non-zero occur. In particular, this implies that the Kac labels of the
representation $\kappa=(k,k')$ will both be odd. 
Since  
$\lim_\clim h_{rr'}=(r-r')^2/4$, 
in the limit \clim\
all the weights $h_\kappa$ of the primary boundary fields will be
integers, and hence all the states in $\cH_{\al\al}$ will have
integer weight.  
If we let $\nH k, k=0,1,\ldots$ be the 
subspace of $\cH_{\al\al}$ of $L_0$ eigenvalue $k$, i.e. the space of
all fields of weight $k$, then
\be
\cH_{\al\al}
= \nH 0 \oplus \nH 1 \oplus \nH 2 \oplus \dots
\label{H-grading}
\ee
Each state in \eref{H-grading} corresponds to a field on the
boundary, and those in \nH 0\ of weight 0 are of special interest.
If $\vec{\psi_i}$ are states of weight 0, and $\psi_i(z)$ the
corresponding field, then the action of the Virasoro algebra
on such fields is%
\footnote{%
In a non-unitary theory, $L_0$ need not be diagonalisable, in which
case equation (\ref{eq:L0comm}) may be replaced by 
$ 
  [ L_m , \psi_i(z) ] 
= z^{m+1}\,{\D\psi_i}/{\D z}
+ z^m (m+1)\, H_{ij}\,\psi_j(z)
\;,
$
where $H_{ij}$ is a nilpotent matrix.
However, by our construction $H_{ij}$ is identically zero.
}
\be
  [ L_m , \psi_i(z) ] 
= z^{m+1}\,\frac{\D\psi_i}{\D z}
\;,
\label{eq:L0comm}
\ee
In a unitary theory, the field $\D\psi_i/\D z$ corresponds to the
state $L_{-1}\vac$ which is a null state, and so can be set to zero in
all correlation functions.
Hence, in a unitary theory, we have 
\be
  [ L_m , \psi_i(z) ] 
= 0
\;,
\label{Lpsi}
\ee
for all fields of weight 0.
Since we can obtain the states in \nH 0\ through a sequence of unitary
models, we expect that the fields will obey \eref{Lpsi} in the limit
\clim. 
(Note that (\ref{Lpsi}) need not always be true for fields of weight 0 --
a counter example is percolation where fields of weight 0 can have
non-trivial space dependence for both $c=0$ \cite{Cardy,Watts} and
$c=1/2$ \cite{StAubin}.) 

The fact that the fields $\psi_i$ are scalars under local conformal
transformations means that they preserve the grading in
\eref{H-grading}. 
In other words, the \OPE\ of such a scalar field with a primary field
of weight $h$ is again a primary field of weight $h$.
However, while the fields of weight 0 are scalars under local
conformal transformations, such transformations cannot alter the order
of fields along a boundary. This means that the fields of weight 0 do
not have to commute with the boundary fields.
To be explicit, if $\psi_i$ are the fields in $\nH 0$,
and those in \nH k\ are denoted by $\Psi_j$,
their \OPES\ take the form
\be
\begin{array}{rll}
   \psi_i(x) \,
   \Psi_j(0)
&= \sum_k C_{ij}^{(+)}{}^k \, \Psi_k(0)
\;,\;\;
&  x>0
\;,
\\[2mm]
   \Psi_j(0) \,
   \psi_i(x) 
&= \sum_k C_{ij}^{(-)}{}^k \, \Psi_k(0)
\;,\;\;
&  x<0
\;,
\end{array}
\label{c<1}
\ee
where $C_{ij}^{(+)}{}^k$ and $C_{ij}^{(-)}{}^k$ need not be equal.
For the A--type models, the couplings between {\em primary} boundary
fields are symmetric as shown in \cite{Runk1}, but for $k>0$, by no
means all the fields in \nH k arise as the limits of primary fields. 

In the particular case of \nH 0, however, all these fields arise as the
limits of primary fields for \clt, so that in this case
 $C^{(+)}_{ij}{}^k = C^{(-)}_{ij}{}^k$
and the fields of weight 0 do commute amongst themselves.
Hence the operator product algebra of the weight 0 fields simplifies
to a straightforward finite-dimensional, commutative, associative algebra:
\[
   \psi_i\, \cdot \,\psi_j
=  b_{ij}^k\, \psi_k
\;.
\]
The space $\cH_0$ consists precisely of (the limits of) all primary
boundary fields with Kac labels $(r,r)$ such that
$N_{(a,a')(a,a')}^{(r,r)}\neq0$, 
i.e. 
\be
  \nH 0 
= \{\; \f_{rr}\; |\; r=1,3,5,\dots,2\min(a,a')-1 \;\} 
\;.
\ee
Since we take all fields on a given boundary $\alpha$ to have unit
norm (\ref{mm:norm}), 
the numbers $b_{ij}^k$ are the limits of the structure constants in
the \OPE\ of three $\f_{rr}$ fields
\be
  b_{ij}^k 
= \lim_\elim \bc {\al}{\al}{\al}{(ii)}{(jj)}{(kk)}
\;,\;\;\;\;
  b_{ij}^1
= \lim_\elim \bc {\al}{\al}{\al}{(ii)}{(jj)}{(11)}
= \delta_{ij}
\;.
\label{eq:bdef}
\ee
It turns out that the finite dimensional associative, commutative
algebra $B$ defined by the constants $b_{ij}^k$ allows a
representation in terms of orthogonal projectors. That is, there are
exactly $\dim B$ elements $P_i\in B$ such that 
$P_i P_j = \delta_{ij} P_i$. 
(The only way this could not be the case would be if there were some
nilpotent elements in $B$, but our choice of normalisation
$b_{ii}^1 = 1$ in eqn.\ (\ref{eq:bdef}) excludes this possibility.)

We have already seen in equation \eref{c<1} that a field
$\f_{rr}\in\nH 0$ acts on the fields in \nH k in different ways for
$x<0$ and $x>0$.
As a result, we have two commuting actions of $B$ on the Hilbert space
$\hat\cH_{\al\al}$, or by an abuse of notation, we have two commuting actions
of two copies $B^R$ and $B^L$ of the algebra, defined by 
\be
\begin{array}{rl}
  \f_{rr} \in B^R 
& :\; \Psi(0) 
  \longmapsto  
  \Psi(0)\, 
  \cdot\, 
  \f_{rr}(-1) 
\;,
\\[1mm]
  \f_{rr} \in B^L 
& :\; 
  \Psi(0) 
  \longmapsto 
  \f_{rr}(1)\, 
  \cdot\,
  \Psi(0) 
\;.
\end{array}
\ee
Given the decomposition (\ref{eq:z}),
\bea
   \lim_\clim Z_{\al\al}
&=& { \sum_{\{\al_i\},\{\al_j\}}} \, \hat Z_{(\hat\al_i)(\hat\al_j)}
\;,
\nn
\eea
it is natural to assume that the projectors $P^L_i$ and $P^R_j$
project onto the subspace $\hat\cH_{\al_i\al_j}$ of $\hat\cH_{\al\al}$ 
corresponding to the 
fundamental boundary conditions $(\hat\al_i)$ and $(\hat\al_j)$ on
either side of the origin, and leads directly to our conjecture 2. 
Again, we have not been able to prove this conjecture, but have
checked it quite extensively, and present the results for the case  
$\al=(2,2)$ in section \ref{sec:22} and for the $(2,p)$ and $(3,p)$
boundaries in sections \ref{app:2p} and \ref{app:3p}.

\subsection{Example: the \clim\ limit of the (2,2) boundary}
\label{sec:22}

In this example we will investigate the spaces 
\nH 0\ and \nH 1\ of the $(2,2)$ boundary, and show how the fields can
be expressed through fields in the superposition of the $(\hat 1)$ and
$(\hat 3)$ b.c.'s. 

For $c$ sufficiently close (but not equal) to one the Hilbert space
of the (2,2) boundary decomposes as
\be
  \cH_{(22)(22)}
= R_{(1,1)} 
\;\oplus\; 
  R_{(3,3)}
\;\oplus\; 
  R_{(1,3)}
\;\oplus\; 
  R_{(3,1)}
\;,
\label{22}
\ee
For \clt\, there are four boundary primary fields, 
  \newcommand{\ph}{\phi}
  \newcommand{\ps}{\psi}
  \newcommand{\pb}{{\bar\psi}}
\be
  1 \equiv \bp{(2,2)}{(2,2)}{(1,1)}
\;,\;\;
  \ph \equiv \bp{(2,2)}{(2,2)}{(3,3)}
\;,\;\;
  \ps \equiv \bp{(2,2)}{(2,2)}{(1,3)}
\;,\;\;
  \pb \equiv \bp{(2,2)}{(2,2)}{(3,1)}
\;,
\label{eq:22fs}
\ee
and their weights are
\be
  h_{1,1} = 0
\;,\;\;
  h_{3,3} = \frac{2 \eps^2}{1{-}\eps}
\;,\;\;
  h_{1,3} = 1{-}2\eps
\;,\;\;
  h_{3,1} = \frac{1{+}\eps}{1{-}\eps}
\;.
\label{eq:22wts}
\ee
In the limit \clim, 
the subspace \nH 0\ is spanned by the two primary fields
\be
  \nH 0 
= \{ 1, \ph \}
\;.
\ee
The only nontrivial \OPE\ amongst these fields is
given in \eref{eq:ph-opes}
\be
   \ph(x) \;\; \ph(y) 
=  1
 \;+\; 
    \fract{2}{\sqrt{3}}\, \ph(y) 
\;,
  \label{eq:33-ope}
\ee
and this defines our algebra $B$, with two generators $1$
(serving as identity) and $\ph$ with relation \eref{eq:33-ope}. 
One identifies the two projectors as
\be
\ba{l}
  P_a = \tfrac 14
        \left( 1 + {\sqrt{3}}\,\ph \right)
\\[2mm]
  P_b = \tfrac 14
        \left( 3 - {\sqrt{3}}\,\ph \right)
\ea
\;,\;\;\;\;
\ba{l}
  1 = P_a + P_b
\\[3mm]
  \ph = \sqrt 3 P_a - \frac 1{\sqrt 3} P_b
\ea
\;.
\label{projector-def}
\ee
To decide which of $P_a$ and $P_b$ is the projector onto the 
$(\hat 1)$ b.c.\ and which onto the $(\hat 3)$ b.c., 
we need the action of these projectors on the weight one fields.

In the limit \clim, the space \nH 1\ is generated by three
primary fields of which 
two are the primary fields $\ps$ and $\pb$ 
corresponding to the spaces $R_{(13)}$ and
$R_{(31)}$ in the decomposition (\ref{22}),
and the third is the field $d_3$ introduced in section
\ref{sec:irreps},
\be
  d_3
\equiv
  \lim_\elim\;
  \frac{\sqrt{1{-}\eps}}{2\eps}\,
  \frac{\D\ph}{\D z}
\;.
\label{eq:22fs'}
\ee
The \OPES\ of the field $\ph$ with the fields of weight one are given
in \eref{eq:ph-opes}, 
from which we can read off
the actions of $\ph(1)=\ph^L$ and
$\ph(-1)=\ph^R$ on the states
$(\vec{\ps} , \vec{\pb} , \vec{d_3})$ 
and assemble them into matrices:
\be
  \ph^L
= \pmatrix{
      0               &  \sqrt{\tfrac 13} & -\sqrt{\tfrac 23} \cr
     \sqrt{\tfrac 13} &  0                & -\sqrt{\tfrac 23} \cr
    -\sqrt{\tfrac 23} & -\sqrt{\tfrac 23} &  \sqrt{\tfrac 13}
  }
\;,\;\;\;\;
  \ph^R
= \pmatrix{
     0                &  \sqrt{\tfrac 13} &  \sqrt{\tfrac 23} \cr
     \sqrt{\tfrac 13} &  0                &  \sqrt{\tfrac 23} \cr
     \sqrt{\tfrac 23} &  \sqrt{\tfrac 23} &  \sqrt{\tfrac 13}
  }
\;.
\ee

We can now determine the action of the four projectors
$P_a^R$, $P_a^L$, $P_b^R$, $P_b^L$ on \nH 1. Of particular interest is the
product of a left and a right projector as it gives the decomposition
\eref{H-decomp}. The result is summed up in the following table:
\vspace{5pt}
{\renewcommand{\arraystretch}{1.6}
\begin{center}
\begin{tabular}{llll}
\hline
  projectors
& image in \nH 1\ 
& \multicolumn{2}{l}{interpretation on $(\hat 1)\oplus(\hat 3)$ boundary}
\\
\hline

  $P_a^L P_a^R$ 
& $0$ 
& no weight 1 field on $\bfb{\hat1}{}{\hat1}$ 
\\
  $P_a^L P_b^R$ 
& $\lambda\cdot(1, ~1,-\sqrt{2})$ 
& $\bfb{\hat 1~~~}{\bph 133}{~~~\hat 3}$ boundary changing field 
\\
  $P_b^L P_a^R$ 
& $\lambda\cdot(1, ~1,~\sqrt{2})$ 
& $\bfb{\hat 3~~~}{\bph 313}{~~~\hat 1}$ boundary changing field 
\\
  $P_b^L P_b^R$ 
& $\lambda\cdot(1,-1,0)$ 
& $\bfb{\hat 3~~~}{\bph 333}{~~~\hat 3}$ boundary field 
\\
\hline
\multicolumn{3}{c}{Table 1:
The images of the projectors on the 
\clim\ limit of the $(2,2)$ boundary}
\end{tabular}
\end{center}
}
\noindent
We see that the interpretation of $P_i$ as projectors is consistent
with the interpretation of the \clim\ limit of the $(2,2)$ boundary 
as $(\hat 1)\oplus(\hat 3)$.
On this superposition, there are six primary boundary fields:
there are two of weight zero (the identity fields $\Id^{(1)}$,
$\Id^{(3)}$ on each boundary condition), three of weight
one (the boundary field $\bph 333$ and the boundary changing 
fields $\bph 133$, $\bph 313$) and one of weight 4
(the boundary field $\bph 335$).
We shall only pay attention to the fields of weight zero and one.
Combining equation \eref{projector-def} and table 1,
we expect the relation between the
$c=1$ fields and the \clim\ limit of $(2,2)$ boundary fields to be
{\renewcommand{\arraystretch}{1.6}
\be
\ba{c}
 \Id^{(1)}
= \frac{1}{4}(1 + \sqrt 3 \ph)
\;,\;\;\;\;
  \Id^{(3)}
= \frac{1}{4}(3 - \sqrt 3 \ph)
\;,
\\

   \bph 333 
=  \lambda_1 \, (\ps - \pb) 
\;,\;\;\;\;
    \bph 133 
= \lambda_2 \, (\ps + \pb - \sqrt{2}d_3) 
\;,\;\;\;\;
    \bph 313 
= \lambda_3 \, (\ps + \pb + \sqrt{2}d_3)
\;.

\ea
\label{field-correspondence}
\ee}%
\noindent%
for some values of $\lambda_1$, $\lambda_2$, $\lambda_3$.

We have already checked that the \OPES\ of the weight zero fields, and
of weight zero fields with weight one fields are in agreement with
this assignment.
Now we check that the 
\OPES\ of the weight one fields on the \clim\ limit of the 
$(2,2)$--boundary, summed up in equation \eref{eq:wt1}, 
reproduces those on the boundary condition
${(\hat1)}{\oplus}{(\hat3)}$ of the \ceq\ model. 
The \OPES\ of the \ceq\ fields can be obtained from the structure
constants given in the appendix.
The non-zero \OPES\ are (all with $x>y$)
{\renewcommand{\arraystretch}{2}%
\be
\begin{array}{c@{\;\;}c@{\;}c@{\;\;}c@{\;\;}c@{\;\;}l@{\;\;}c@{\;\;}l}

    \Id^{(a)} 
&   \Id^{(b)} 
&=& \delta_{a,b} \; \Id^{(a)}
\;,
\\

    \Id^{(a)} 
&   \bph bc3(y)
&=& \delta_{a,b} \; \bph bc3(y)
\;,
\\

    \bph bc3(x)
&   \Id^{(a)} 
&=& \delta_{a,c} \; \bph bc3(x)
\;,
\\

    \bph 133(x)
&   \bph 313(y) 
&=& \frac 1{(x{-}y)^2}\, \Id^{(1)} 
&+& O(1)
\;,
\\

    \bph 133(x)
&   \bph 333(y) 
&=& ~\frac 2{(x{-}y)}\, \bph 133(y) 
&+& 
   O(1)
\;,
\\

    \bph 313(x)
&   \bph 133(y) 
&=& \frac 13\,\frac 1{(x{-}y)^{2}}\, \Id^{(3)} 
&+& \frac 23\,\frac 1{(x{-}y)} \,  \bph 333(y) 
&+& O(1)
\;,
\\
   
    \bph 333(x)
&   \bph 313(y) 
&=& ~\frac 2{(x{-}y)}\, \bph 313(y) 
&+& O(1)
\;,
\\

    \bph 333(x) 
&   \bph 333(y) 
&=& \frac 1{(x{-}y)^{2}}\, \Id^{(3)} 
&+& \frac 1{(x{-}y)} \,  \bph 333(y) 
&+& O(1)
\;.
\end{array}
  \label{c=1:1+3}
\ee
}%
The \OPES\ such as 
$\bph 133(x) \bph 133(y)\,$,  $\,\bph 313(x) \bph 313(y)$,
etc, have to be 
zero because the boundary conditions do not match up. 
Substituting (\ref{field-correspondence}) one verifies that these
\OPES\ vanish for any choice of
$\lambda_1,\lambda_2,\lambda_3$,
and that the five nontrivial \OPES\ in
(\ref{c=1:1+3}) are correctly reproduced for
$\lambda_1=-\sqrt{3/8}$ and $\lambda_2\lambda_3 = 1/{16}$.

This does not, of course, represent a complete proof of our
conjecture, since we have not treated the weight four primary boundary
field, nor any fields of weight greater than one, and we do not in any
case have an independent proof that the \ceq\ b.c.'s $(\hat r)$ arise
as the boundary conditions of any particular \ceq\ bulk theory, but we
regard it as very strong evidence for conjectures 1 and 2.

\newpage

\resection{The perturbations by $\f_{rr}$}
\label{sec:c<1}

We can now examine the $\f_{rr}$ perturbations of the minimal models
in the light of these results.

The first point is that the situation at \ceq\ is entirely clear.
Consider the \clim\ limit of the field $\f_{rr}$ on the $(a,a')$
boundary condition. This can be expanded in terms of the projectors
$P_i$ onto the fundamental components on the decomposition
(\ref{eq:conj})
\be
  \f_{rr} 
= \sum\, 
  \mu^{(r)}_i \, P_i
\;.
\label{eq:frr}
\ee
We consider the perturbation of the theory on a strip of width $R$
with boundary condition $\al = (a,a')$ on the right edge and 
$\beta = (b,b')$ on the left by the addition of the field 
$\f_{rr}^{(\al\al)}$ on the right edge, 
\be
  S 
= S_0 \;+\; \lambda \int \f_{rr}^{(\al\al)}(x)\,\D x
\;.
\label{eq:dS}
\ee
This can be reformulated in terms of a perturbed Hamiltonian on the UHP
\bea
  H
&=&
  \left(\frac{\pi}{R} \right)
  \, \left[\,
  L_0 \,-\, \frac{c}{24} 
  \;+\;
  \lambda\,\f_{rr}(1)
  \right]
\;=\;
  \left(\frac{\pi}{R} \right)
  \, \left[\,
  L_0 \,-\, \frac{c}{24} 
  \;+\;
  \lambda\,\sum\,\mu^{(r)}_i\,P^R_i
  \right]
\;.
\label{eq:H}
\eea
It is clear that the effect of the addition of the perturbation is
just to add an amount $\lambda\,\mu^{(r)}_i$ to the energy of a
state in the fundamental component $i$ on the right boundary.
As $|\lambda|\to\infty$, only the sector(s) with minimal
$(\lambda\mu^{(r)}_i)$ will survive (with all the other sectors
decoupling) i.e. for $\lambda>0$, the right boundary flows to the
system with  
boundary condition $\oplus(\hat c)$ with $\mu^{(r)}_{c}$ minimal, and 
for $\lambda<0$, the right boundary flows to the system with
boundary condition $\oplus(\hat c)$ with $\mu^{(r)}_{c}$ maximal.

To return to the example we have treated in depth, consider the model
with boundary condition $(11)\equiv(\hat 1)$ on the left edge and
$(22)\equiv(\hat 1)\oplus(\hat 3)$ on the right edge.
There is a single non-trivial boundary field \ftt\ on the
$(22)$ boundary, which can be expressed in terms of the projectors
$P_{(\hat 1)}$ and $P_{(\hat 3)}$ as
\be
  \ftt
= \sqrt 3\,P_{(\hat 1)} \;-\; \frac{1}{\sqrt 3}\,P_{(\hat 3)}
\;.
\label{eq:f33}
\ee
So, for $\lambda>0$ this system flows to the model on 
the strip with boundary conditions $(\hat 1)$ on the left and 
$(\hat 3)$ on the right, and for $\lambda<0$ it flows to the model with
boundary condition $(\hat 1)$ on the left and $(\hat 1)$ on the right.

There are several ways we can present this graphically.
For a general perturbation, $\lambda$ is not a
dimensionless variable, so we define the
dimensionless variables $\kappa$ and $r$,
\be
  \kappa
= \lambda\, R^{y}
\;,\;\;
  r 
= |\kappa|^{1/y}
\;,\;\;\;\;
  y 
= 1 - h_{33}
\;.
\label{eq:kappadef}
\ee
In figure \ref{1}a we can plot the eigenvalues of $(R/\pi)H$ against
$\kappa$ for fixed $R$, and in figures \ref{1}b and \ref{1}c we
plot the gaps $R(E - E_0)/\pi$ above the ground state energy against
$\log|r|$ for $\lambda$ positive and negative respectively.
We show these for later comparison with the equivalent plots for \clt.

We should comment that the apparent lack of smoothness 
in figure 1b at $\kappa=\sqrt 3/4$ is due to the fact we are plotting
the scaled energy {\em gaps}, and  that the first excited state
crosses the ground state at that value of $\kappa$.

\[
\begin{array}{lll}
\refstepcounter{figure}
\label{1}
\epsfysize 5.5truecm
\epsfbox{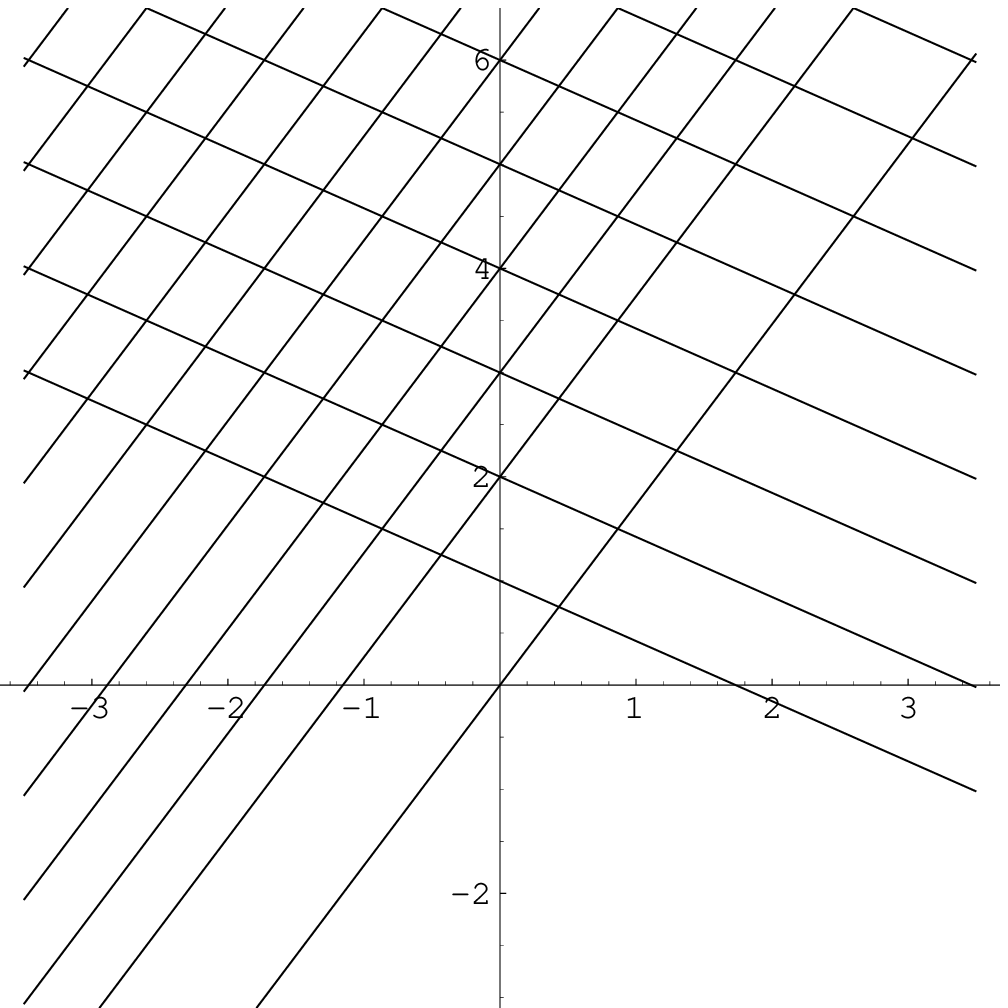}
&
\epsfysize 5.5truecm
\epsfbox{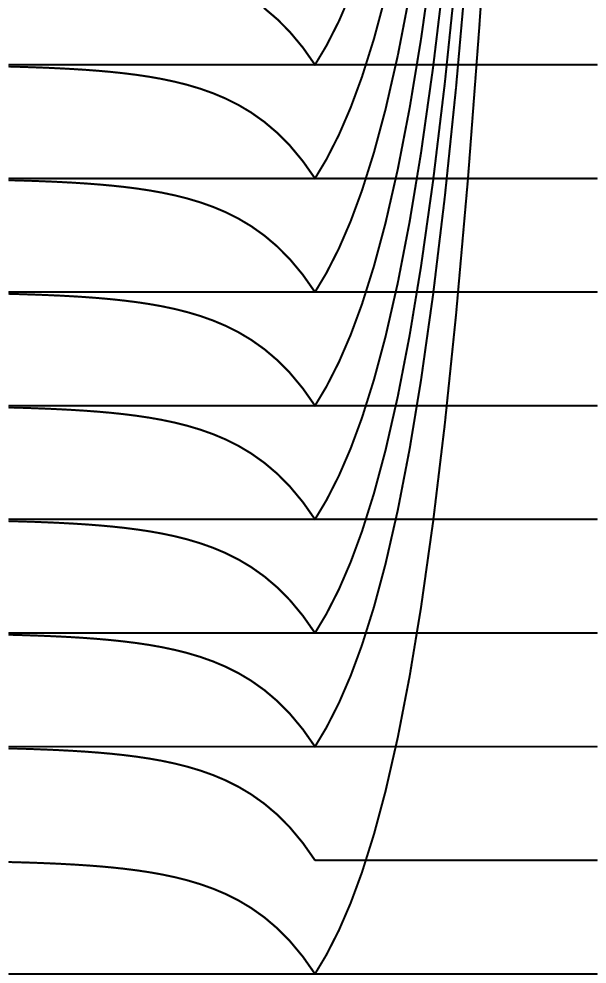}
&
\epsfysize 5.5truecm
\epsfbox{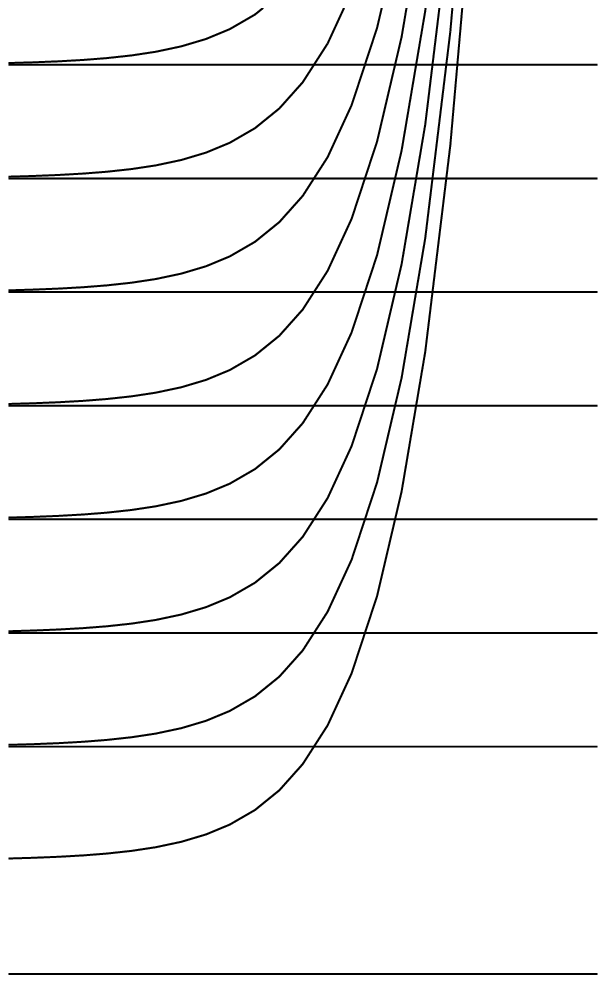}
\\
\parbox{5 truecm}{
\small\raggedright
Figure \ref{1}a:
$(22) \pm \lambda\ftt$.
\\
The eigenvalues of $(R/\pi)H(\kappa)$ plotted against $\kappa$.
}
 &
\parbox{4 truecm}{
\small\raggedright
Figure \ref{1}b:
$(22) + \lambda\ftt$.
\\
Energy gaps plotted against $\log|r|$.
}
 &
\parbox{4 truecm}{
\small\raggedright
Figure \ref{1}c:
$(22) - \lambda\ftt$.
\\
Energy gaps plotted against $\log|r|$.
}
\end{array}
\]

\subsection{The minimal models with \clt}
\label{sec:c<1b}

For \clt\  but $\epsilon$ still small, we expect that this picture will
only change slightly, and in particular the IR end points of these
flows should agree with the results at \ceq.
This leaves some ambiguity, however, since several different boundary
conditions for \clt\  may have the same limit at \ceq.
In the particular case of the $(22)$ boundary condition perturbed by
\ftt, the IR limits at \ceq\ are $(\hat 1)$ and $(\hat 3)$ for
$\lambda$ negative and positive respectively.
There is a single boundary condition which has as its limit
$(\hat 1)$, namely $(11)$; however both $(13)$ and $(31)$ have as
their limits $(\hat 3)$. We must find a way to decide which is the
correct IR endpoint for \clt.

One method might be to use conformal perturbation theory, but since
the perturbation is UV--finite and IR--divergent, the conformal
perturbation theory tells us nothing about the IR end point, as was
also the case for the Lee-Yang model studied in \cite{DPTWa1,DRTWa1}.

\begin{figure}
\[
\begin{array}{lll}
\refstepcounter{figure}
\label{2}
\epsfysize 5.201truecm
\epsfbox{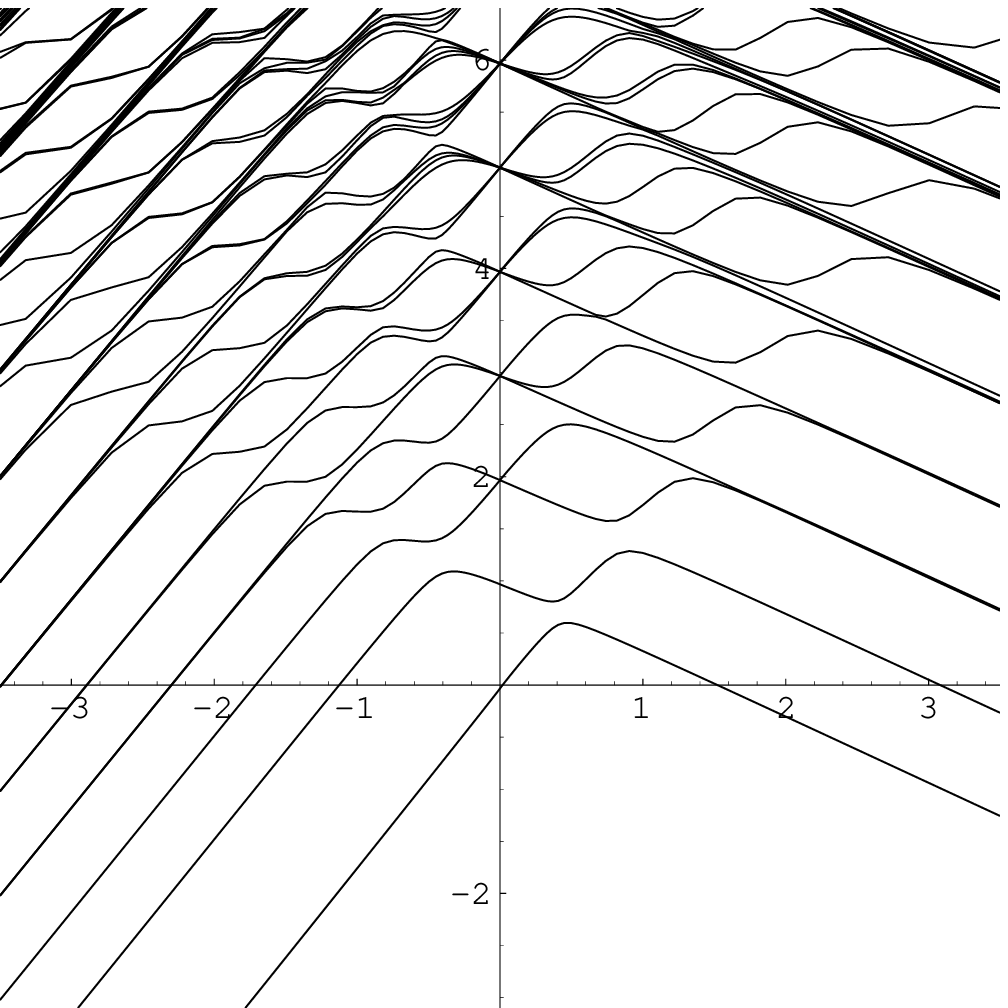}
&
\epsfysize 5.201truecm
\epsfbox{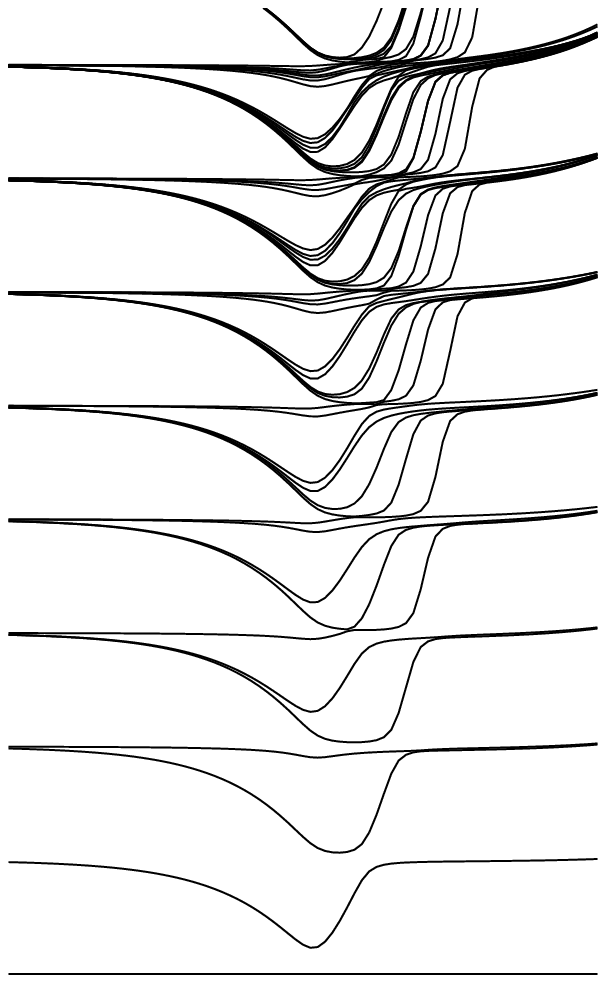}
&
\epsfysize 5.201truecm
\epsfbox{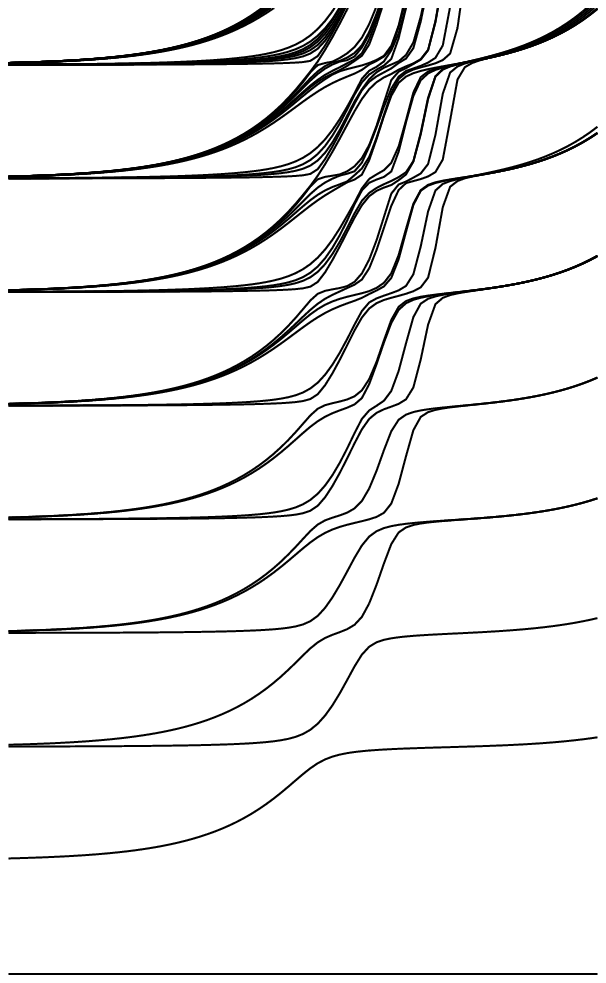}
\\
\parbox[t]{5 truecm}{
\small\raggedright
Figure \ref{2}a:
$M_{10,11}: (22) \pm \lambda\ftt$.
}
 &
\parbox[t]{4 truecm}{
\small\raggedright
Figure \ref{2}b:
$M_{10,11}: (22) + \lambda\ftt$.
}
 &
\parbox[t]{4 truecm}{
\small\raggedright
Figure \ref{2}c:
$M_{10,11}: (22) - \lambda\ftt$.
}
\\[6mm]
\refstepcounter{figure}
\label{3}
\epsfysize 5.201truecm
\epsfbox{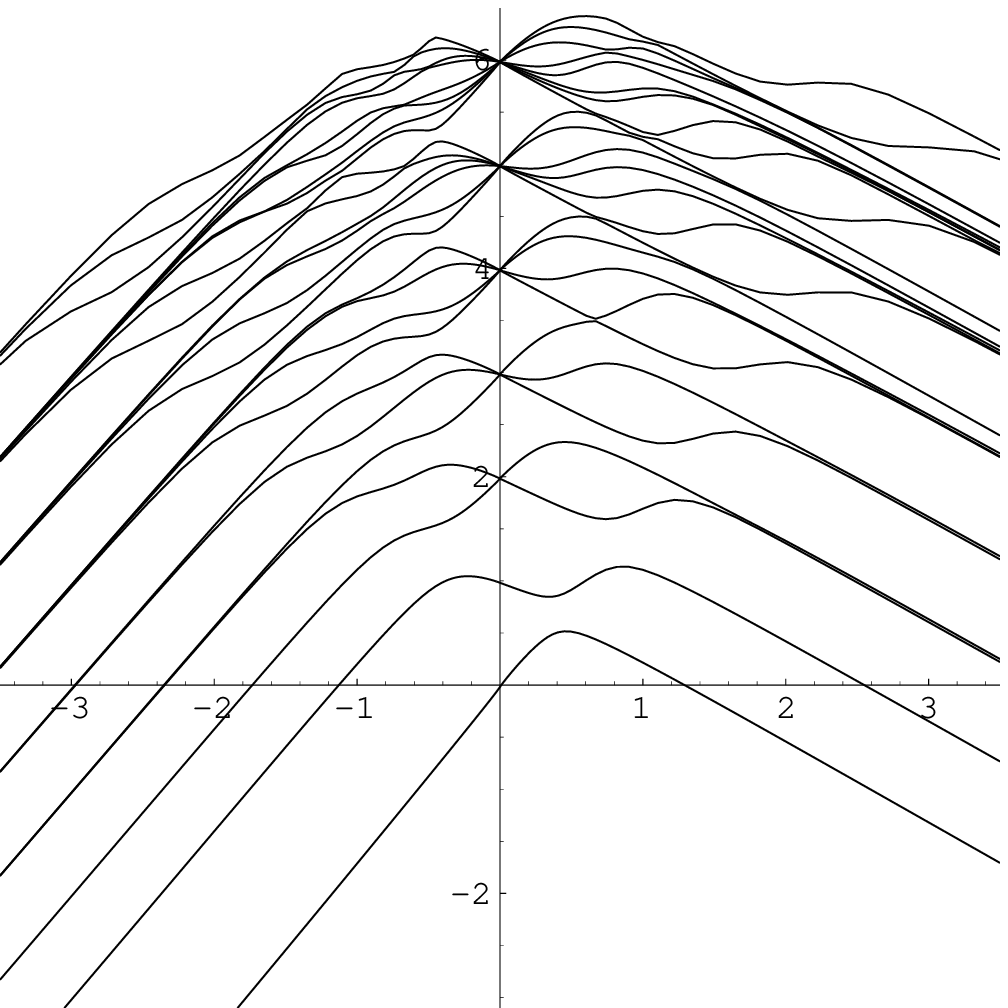}
&
\epsfysize 5.201truecm
\epsfbox{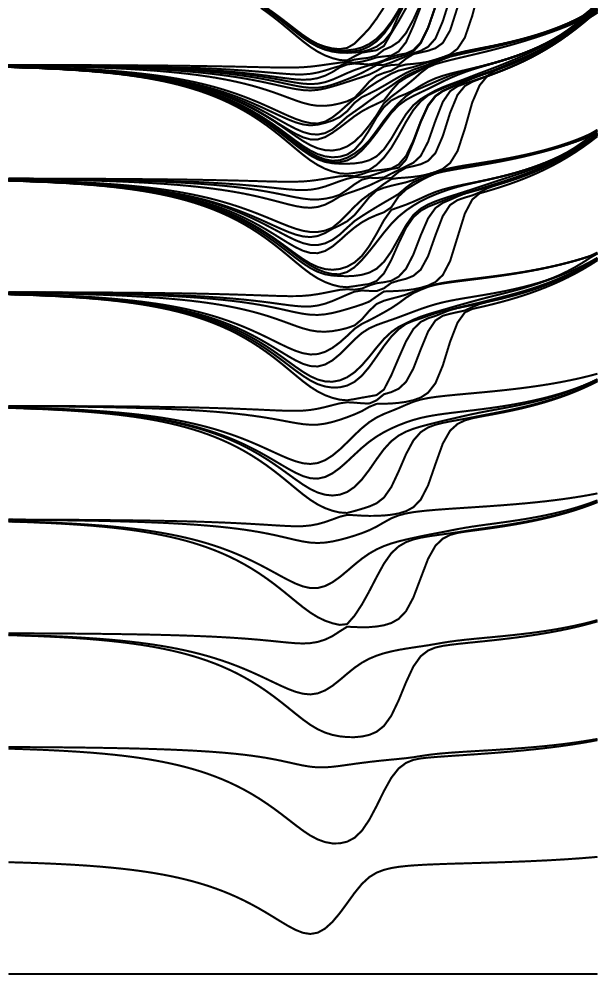}
&
\epsfysize 5.201truecm
\epsfbox{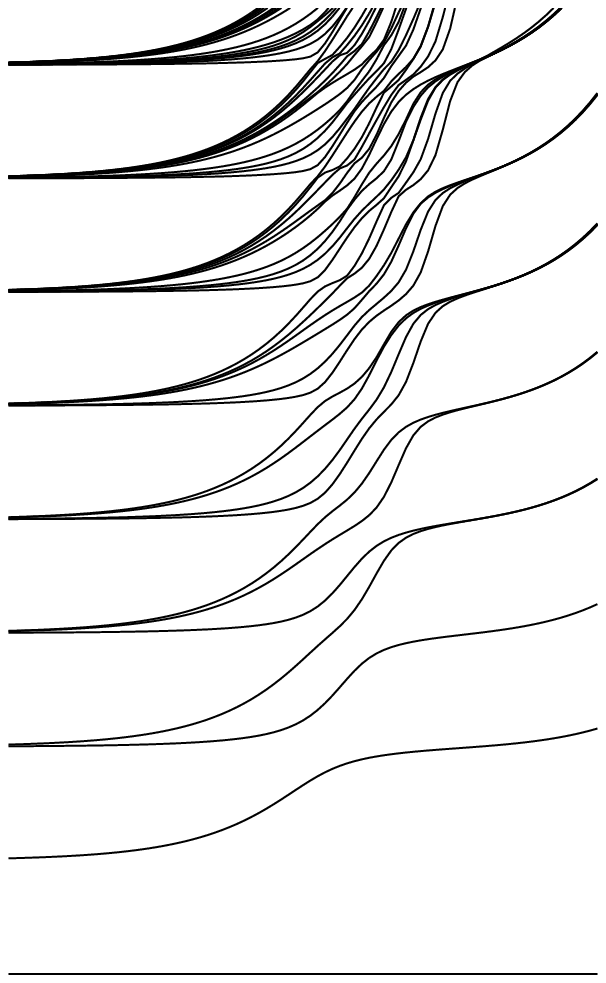}
\\
\parbox[t]{5 truecm}{
\small\raggedright
Figure \ref{3}a:
$M_{6,7}: (22) \pm \lambda\ftt$.
}
 &
\parbox[t]{4 truecm}{
\small\raggedright
Figure \ref{3}b:
$M_{6,7}: (22) + \lambda\ftt$.
}
 &
\parbox[t]{4 truecm}{
\small\raggedright
Figure \ref{3}c:
$M_{6,7}: (22) - \lambda\ftt$.
}
\\[6mm]
\refstepcounter{figure}
\label{4}
\epsfysize 5.201truecm
\epsfbox{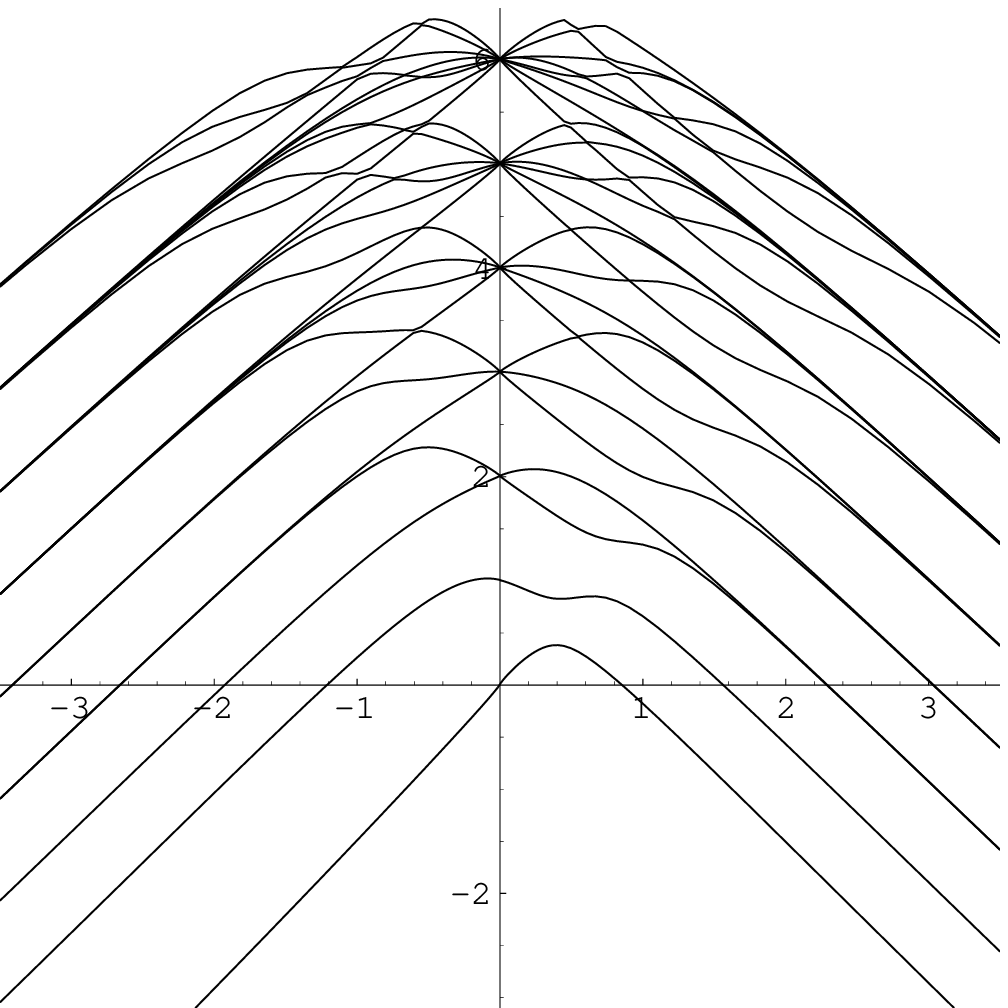}
&
\epsfysize 5.201truecm
\epsfbox{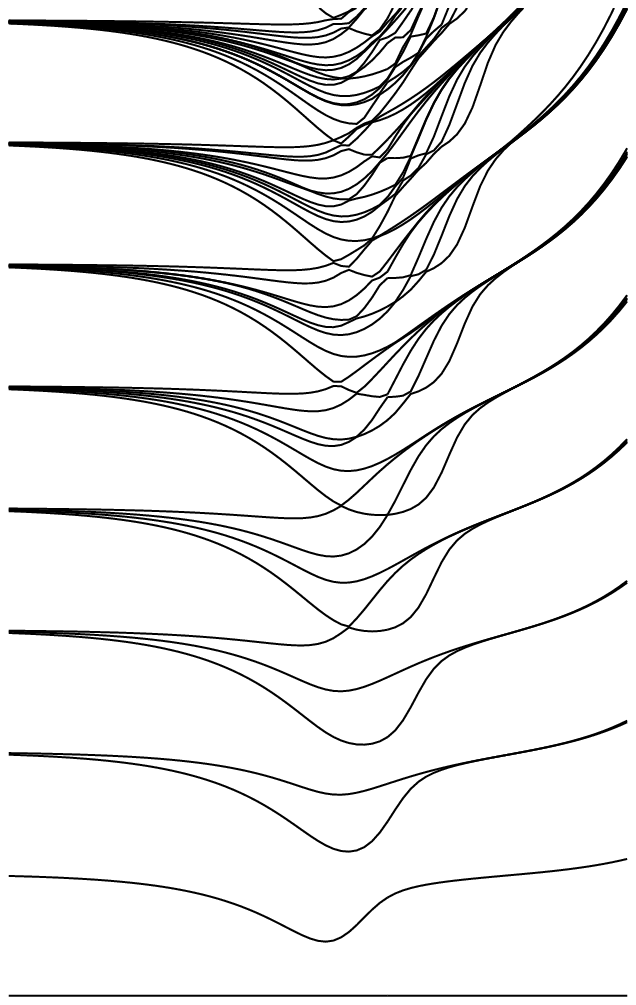}
&
\epsfysize 5.201truecm
\epsfbox{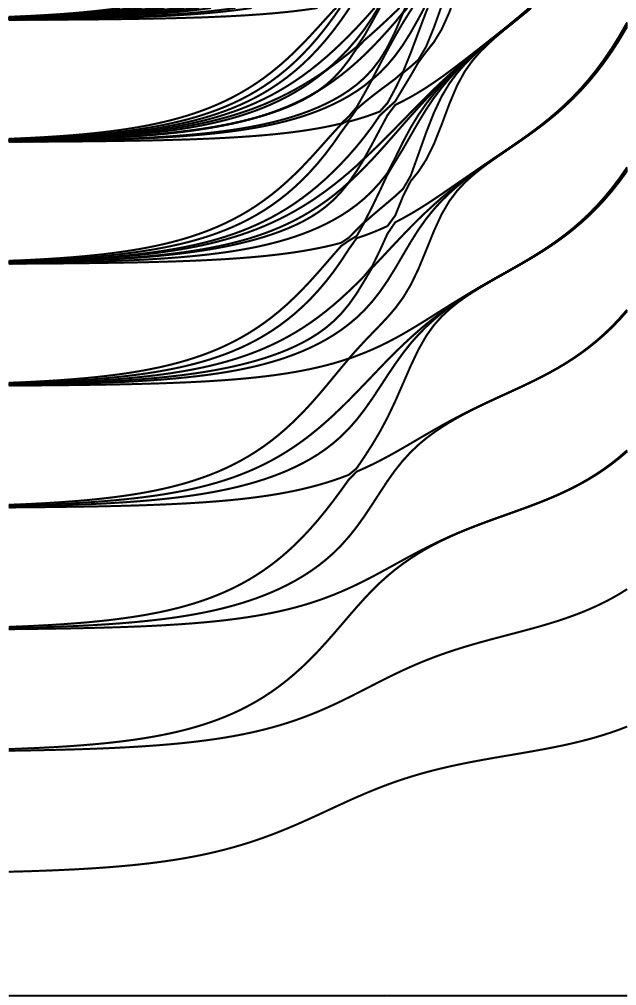}
\\
\parbox[t]{5 truecm}{
\small\raggedright
Figure \ref{4}a:
$M_{4,5}: (22) \pm \lambda\ftt$.
}
 &
\parbox[t]{4 truecm}{
\small\raggedright
Figure \ref{4}b:
$M_{4,5}: (22) + \lambda\ftt$.
}
 &
\parbox[t]{4 truecm}{
\small\raggedright
Figure \ref{4}c:
$M_{4,5}: (22) - \lambda\ftt$.
}
\\
\multicolumn{3}{l}{
\parbox[t]{\linewidth}{\small\raggedright
Figures \ref{2}a, \ref{3}a and \ref{4}a:
The first 25 eigenvalues of $(R/\pi)H(\kappa)$ plotted against
$\kappa$;
\newline
Figures 
\ref{2}b,\ref{2}c,
\ref{3}b,\ref{3}c,
\ref{4}b and \ref{4}c:
Energy gaps $(R/\pi)(E_i{-}E_0)$ plotted against $\log|r|$
}}
\end{array}
\]
\end{figure}

The only other method open to us at the moment is 
the Truncated Conformal Space Approach (TCSA) \cite{DPTWa1}.
In figures \ref{2}a--\ref{4}c we show the equivalent plots for the
perturbation of the strip with boundary conditions $(11)$ and $(22)$
by \ftt\ for the minimal models $M_{10,11}$, $M_{6,7}$ 
and $M_{4,5}$, all calculated using TCSA.

We see that the pattern in $M(10,11)$ is extremely similar to that at
\ceq, but that the lines no longer cross. This is the typical behaviour
of a non-integrable flow -- the folklore being that one can only
expect line crossings 
if there are conserved quantities present which forbid mixing of
states. These `gaps' open up further as $\epsilon$ grows in $M(6,7)$
and $M(4,5)$, but we see that in $M(4,5)$ the lines appear to cross
again. This would indicate that this flow is again integrable with an
infinite number of conserved quantities, and closer inspection shows
that this is indeed likely to be the case as the symmetries of the Kac
table mean that in the model $M(4,5)$, $\ftt\equiv\f_{12}$,
and $\f_{12}$ is well known to be one of the generic integrable
perturbations (along with $(1,3)$ and $(1,5)$ and the images under
$(r,s)\leftrightarrow (s,r)$. )
As shown in \cite{GZ}, the same arguments that are used to show the
integrability of bulk perturbations can also be used to show the
integrability of the analogous boundary perturbations.
We have checked that there are also line-crossings in the perturbations
of the boundary condition $(33)$ by the field
$\f_{55}\equiv\f_{12}$ in the model $M(6,7)$.
and of the boundary condition $(44)$ by the field
$\f_{77}\equiv\f_{12}$ in the model $M(8,9)$.

We should also comment on the fact that the energy gaps do not tend to
constants in the IR. This is due to truncation effects, and the effect
decreases as the truncation level is increased.

The boundary condition $(\gamma)$ at the IR fixed point for positive
$\lambda$ could be deduced from the TCSA plots in one
of two ways.
Firstly, we could use the asymptotic behaviour of the ground state
energy to find the conformal weight of the ground state directly
but this is problematic, as we shall discuss in the next section.

The easier method is to count the degeneracies of the IR spectrum,
from which  one can identify the  partition function of the strip with
boundary conditions $(11)$ and $(\gamma)$, 
since the partition function of a strip with boundary conditions
$(11)$ and $(\gamma) = \oplus (\gamma_i)$ is
equal to the sum of characters of the
representations $\gamma_i$:
\be
  Z_{(11)(\gamma)}
= {\textstyle \sum_i}\;\; \chi^{}_{\gamma_i}
\;.
\ee
In this case, we expect that $(\gamma)$ will be a single
representation $(13)$ or $(31)$, and so the partition function should
be $\chi^{}_{13}(q)$ or $\chi^{}_{31}(q)$ respectively.
We can identify the character by the counting of states, which will
show the existence of null vectors at levels $3$ and $(p-1)(p'-3)$ for
boundary condition $(13)$ and at levels $3$ and $(p-3)(p'-1)$ for
boundary condition $(31)$.
For the models $M_{10,11}$ and $M_{6,7}$, the second null vector is at
too high a level to be calculated easily using TCSA, but for $M_{4,5}$
it should be at level 6 or 4 for the cases (13) and (31) respectively.
From figure \ref{4}b we see that there is indeed a state missing at
level 4, so we can positively identify this IR endpoint as the (31)
representation. 

We should make it clear that we cannot {\em prove} using TCSA that the
IR endpoint of the flow $(22)-\ftt$ is the b.c.\ $(1,1)$, since 
(quite apart from numerical errors) one can never be sure that the IR
regime has been reached. 
At best we can say that for $\log|r|\sim2$ (the right hand edge of the
graphs \ref{2}b, \ref{3}b and \ref{4}b) the counting of states
indicates that the flow is in the neighbourhood of the $(31)$ b.c.,
and since that b.c.\ has no relevant perturbations it is reasonable
to believe that it is the endpoint of the flow.
Similarly, since the counting of states indicates that the flows
$(22)+\ftt$ enter the vicinity of the $(31)$ b.c.~which again has
no relevant perturbations, this suggests that this is indeed the IR
endpoint of this flow. 

\subsection{The scaling behaviour of the ground state energy}

We expect the ground state energy 
calculated using TCSA to have three different scaling behaviours
according to the value of $R$:
\be
  f(r) 
\equiv
 \frac{R E_0(R)}{\pi} 
\;\sim\;
\cases{
  (h_{\hbox{\tiny UV}} - \frac c{24}) \;+\; c_1\, r^y \;+\; \ldots
~~
  & {\small $R$ small}, \cr
  (h_{\hbox{\tiny IR}} - \frac c{24}) \;+\; c_2\, r   \;+\; \ldots
  & {\small The `scaling region',} \cr
                               c_3\, r^y \;+\; \ldots
  & \parbox[t]{5cm}{\small\raggedright%
{$R$ large, truncation errors dominate.}}
}
\label{eq:E0}
\ee
(here $c_2$ is the IR boundary free-energy-per-unit-length, and 
 $h_{\hbox{\tiny UV/IR}}$ are the minimal weights of the UV and IR fixed points
respectively.)

It is well known that the TCSA method cannot be applied easily to
bulk massless flows as it is hard to reach the appropriate scaling
region, for several reasons. Firstly, since the fixed point may still
have relevant perturbations, errors introduced by truncation can drive
the flow away from the intended fixed point, and the
corrections to the leading scaling behaviour can be large,
decaying with powers of $r$, rather than exponentially. 
Secondly, high level states that are dropped by truncation can still
contribute appreciably to the ground-state-energy.
A recent exception to this rule is the double-Sine-Gordon model,
where it has proven possible to obtain the flow to the Ising point using
TCSA \cite{BPTWa1} by truncating at rather high levels and so decreasing the
truncation errors, and by fine-tuning in two variables to hit the IR
fixed point. 
To test for the onset of scaling, we can try to fit the ground state
energy by a function of the form
\be
  f(r)
\sim\;
  a \;+\; c\, r^b
\;,
\label{eq:scaling} 
\ee
and estimate $b$ by the function
\be
  b_{\rm est}(r)
= 1 
+ r \frac{\D}{\D r} 
  \log\Big( \,
  \frac{\D f}{\D r}  \,\Big)
\;.
\ee
In the scaling regime, we should obtain $b\approx 1$.
Similarly, we can estimate $a$ by using the expected scaling form
\eref{eq:E0} (i.e. taking $b=1$ in \eref{eq:scaling}) to give
\be
  a_{\rm est}(r) 
= - \, r^{2} \,
    \frac{\D}{\D r}(\, f/r \,)
\;.
\ee
In figure \ref{fig:scaling} we plot 
$a_{\rm est}(r)$ and $b_{\rm est}(r)$ against $\log|r|$
for the model $M_{6,7}-\lambda\ftt$ calculated using TCSA with
levels 6, 10 and 15, that is truncated to 26, 109 and 489 states
respectively, and also an extrapolation of the data to infinite level.
Also shown in these plots are the expected UV and IR behaviour.
Although the finite level TCSA results do not show scaling --
$b_{\rm est}$ does not tend to 1 and $a_{\rm est}$ does not tend to
the expected constant -- the extrapolated results are much better. 
However,  even after extrapolation, we cannot really say that we have
shown that the IR limit is indeed the one we expect. 

\[
\ba{ll}
\refstepcounter{figure}
\label{fig:scaling}
\epsfxsize .45\linewidth
\epsfbox{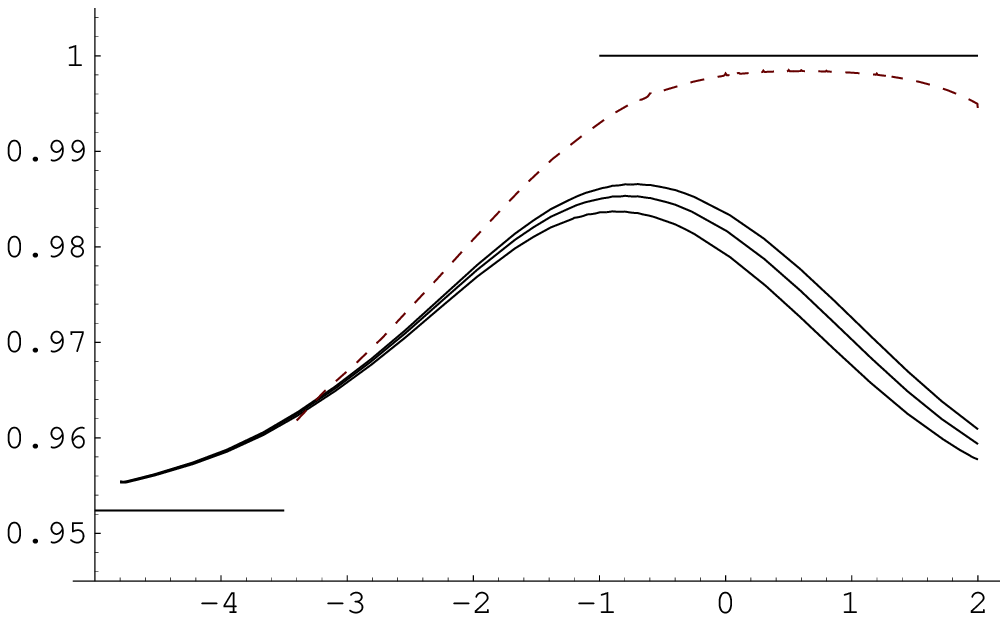}
&
\epsfxsize .45\linewidth
\epsfbox{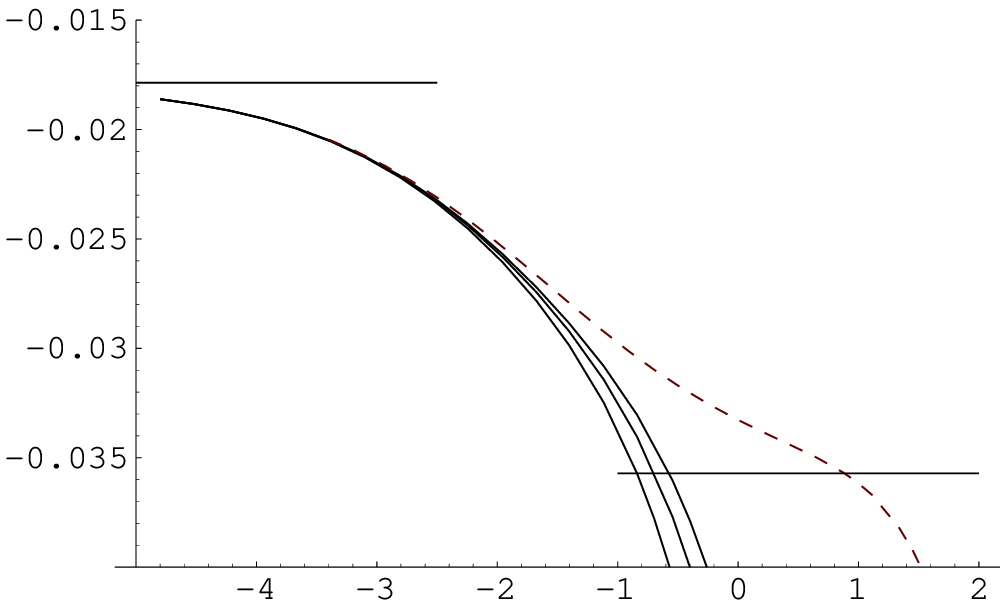}
 \\
\parbox{.4\linewidth}{\small\raggedright
Figure \ref{fig:scaling}a}
&
~~~~
\parbox{.4\linewidth}{\small\raggedright
Figure \ref{fig:scaling}b}
\\
\multicolumn{2}{l}{
\parbox{.9\linewidth}{\small\raggedright
Plots of 
$b_{\rm est}(r)$ (left) and
$a_{\rm est}(r)$ (right) from TCSA for $M_{6,7}: (22) - \lambda\ftt$.
The solid lines are from truncation to levels 6,10 and 15, and the
dashed line an extrapolation to infinite level.
Also shown are the expected UV and IR values (the horizontal lines).
}}
\ea
\]

We see from these plots that the dominant contribution to the errors
comes from the truncation -- the extrapolation of $b_{\rm est}(r)$ in
figure \ref{fig:scaling}a suggests that scaling would set in for 
$\log|r|\,{\gtrsim}\;0$, if the truncation errors could be removed.
Unfortunately the corresponding
results for the flow in the positive direction are not even as good,
since the scaling region only appears to set in for $\log|r|>2$. 

We can try to improve on these results by including in our fits some
of the sub-leading contributions to
\eref{eq:E0}
\be
  f(r)
\;=\;
  a 
\;+\; 
  c\, r 
\;+\;
  \sum d_i\, r^{1-h_i}
\;+\;
  \ldots
\;,
\ee
corresponding to the leading contributions from the quasi-primary irrelevant
operators of weight $h_i$ at the IR fixed point.
These operators are $T(x)$ on the $(11)$ b.c., and $T(x)$ and
$\f_{31}$ on the $(31)$ b.c.
While these do improve the fit to the IR behaviour, 
even including these extra corrections we do not see unambiguous signs
that we are at the correct IR fixed point, and so do not present them here.

\subsection{The nonunitary models}

The question we must ask now, is whether this scheme we have outlined
is also valid for perturbations of the non-unitary models $M(p,p')$
with $p'\neq p+1$. 

For those models far from $t=1$, 
the field $\f_{rr}$ has weight greater than 1/2. 
This means that in a proper field theory treatment the model needs to
be regularised and renormalised and a large range of possible counter
terms need to be considered. 
We certainly have no expectation that our results
(which are based on the idea that $\f_{rr}$ is close to a scalar
field) will remain true in such a case.
However, one might hope that for $p'$ close to $p$ this picture would
still work.
To answer this question it is important first to address the general
dependence of the pattern of the flows on the boundary condition, the
perturbing field and the central charge.

We first consider the {
generic} situation with the
parameter $t$ irrational, and where the boundary condition $(h)$ and
perturbing field $\f_{h'}$ are also {generic}, i.e. for which
there are no null states in the representation $R_h$.
In this case the spectrum depends {
smoothly} on the parameters $t$, $h$
and $h'$
(n.b.~we are not making any assertions about the existence or otherwise of
a local field theory with these properties, only about the TCSA
spectra as determined by the TCSA matrix elements).
The only singularities occur when $t$ and $h$ are such that there are
null states in $R_h$, in which the spectrum is given by the generic
pattern but with the {
omission} of certain complete lines
corresponding to the decoupling of the null states%
\footnote{A similar phenomenon occurs for the bulk perturbations by
the field $\f_{13}$, where the minimal model spectra are a {
subset} of the spectra of the folded sine-Gordon model \cite{BPTWa2}}.%
If the null states that are decoupled are above the truncation level,
then no difference will be seen on the TCSA plots.

\begin{figure}
\[
\begin{array}{lll}
\refstepcounter{figure}
\label{fig:5}
\epsfysize 8.01truecm
\epsfbox{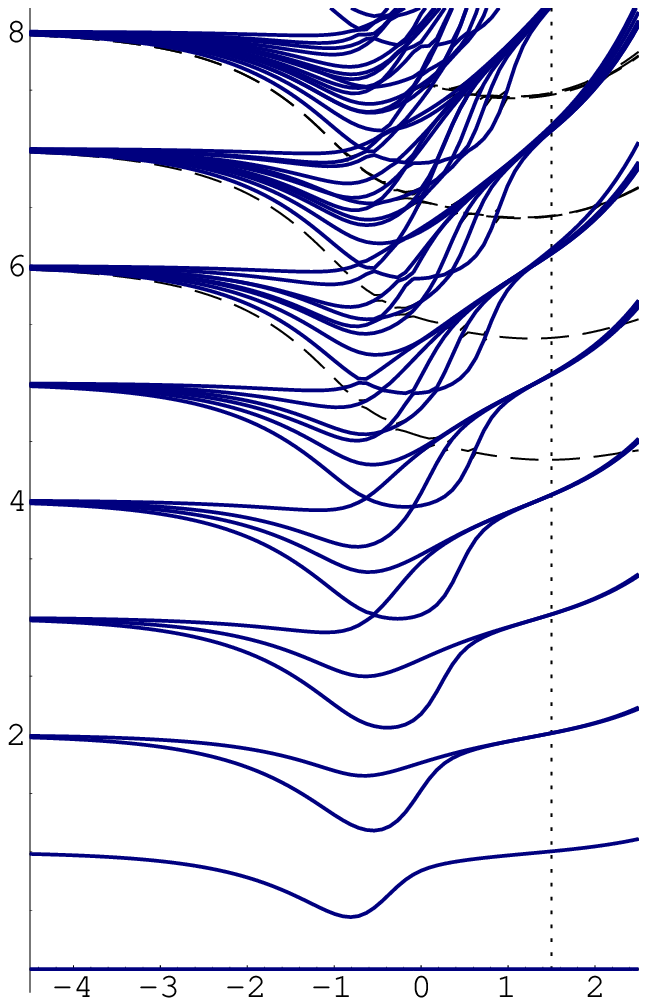}
&~~~~&
\epsfysize 8.01truecm
\epsfbox{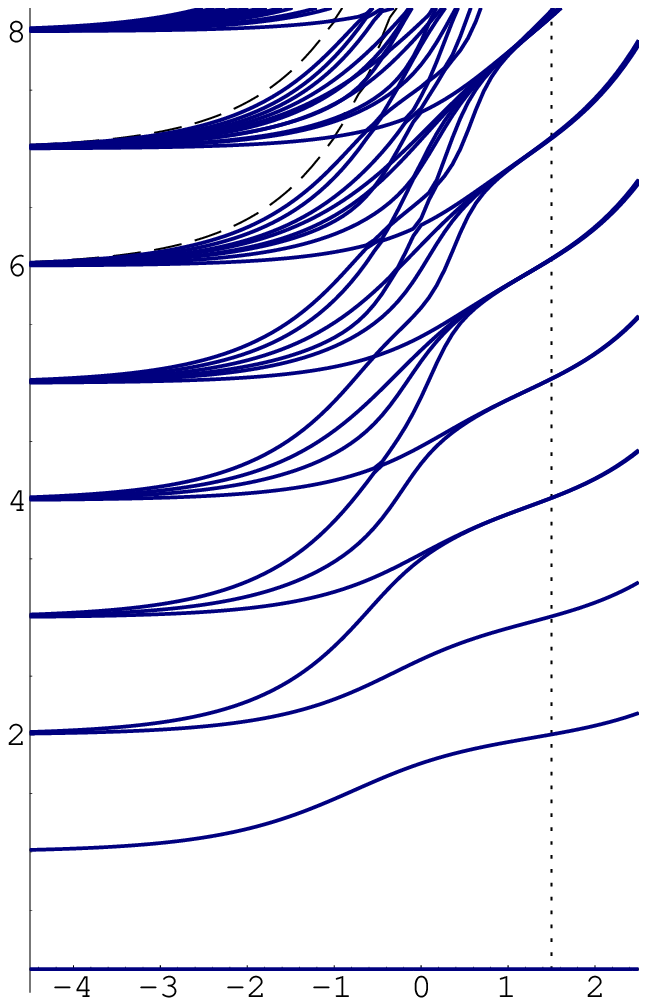}
\\
\parbox{4 truecm}{
\small\raggedright
Figure \ref{fig:5}a}
& &
\parbox{4 truecm}{
\small\raggedright
Figure \ref{fig:5}b}
\\
\multicolumn{3}{l}{
\parbox{11truecm}{\small\raggedright
Energy gaps plotted against $\log|r|$
for $M_{t=0.8002}: (22) + \lambda\ftt$ (left)
and $M_{t=0.8002}: (22) - \lambda\ftt$ (right)
from TCSA to level 14 (369 states).
}}
\end{array}
\]%
\vspace{-0.5cm}
\end{figure}

In this section we have mostly focussed on the flows
$(22)\pm\ftt$, for which there are always null states in the
representation $R_{(2,2)}$, starting at level 4. 
There are {
extra} null states for rational values of the parameter
$t=p/p'$, starting at level $(p-2)(p'-2)$.
In figures \ref{2}a--\ref{4}c, we have shown states up to level 8, so
that the pattern is generic (for these flows) apart from the cases
\be
  t
\; \in \;
  \{\; \frac 34\;,\;\; \frac 35\;,\;\; \frac 37\;,\;\; \frac 38\;,\;\;
       \frac 45 \;\}
\;.
\label{eq:special-ts}
\ee
The last case we have looked at $M(4,5)$, is one of these special
values, so that we should also look at a `neighbouring' flow to see
the generic situation.
In figures \ref{fig:5}a--b we show the plots 
for $t=0.8$ (truncated to level 14) in bold, with the extra lines for
$t=0.8002$ superposed as dashed lines. We also indicate (with a
vertical dotted line) how far we think the qualitative features of
this graph can be trusted. 

\begin{figure}[h]
\[
\begin{array}{c}
\refstepcounter{figure}
\label{fig:6}
\epsfysize 5.201truecm
\epsfbox{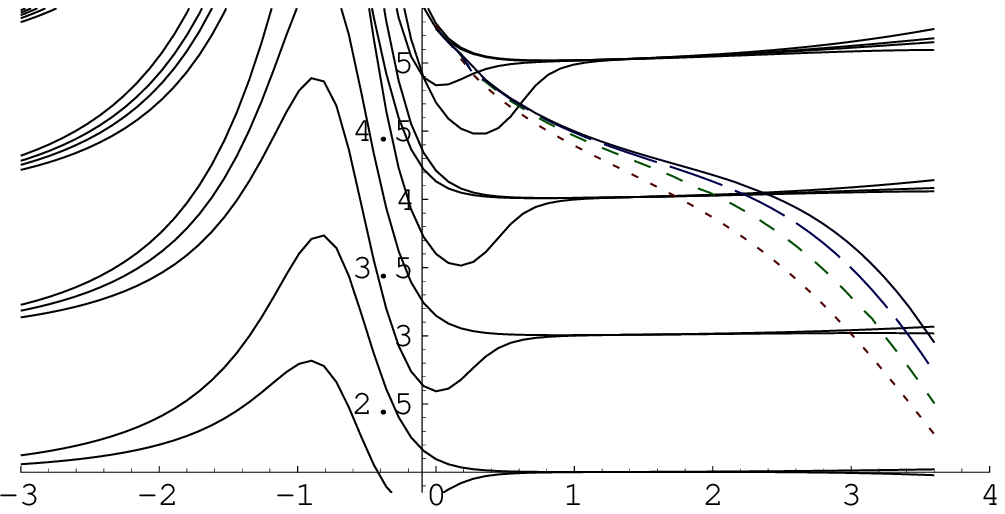}
\\
\parbox{12 truecm}{
\small\raggedright
Figure \ref{fig:6}:
$(22) + \lambda\ftt$.
\\
The normalised energy gaps for $M(4,5)$ plotted against $\log|r|$,
with the first `extra' line for $M(4001,5000)$ superposed.
Truncation levels are 
 8 (dotted), 10 (short dashed), 12 (long dashed) and 14 (solid).
}
\end{array}
\]\vspace{-0.5cm}
\end{figure}

It is important to work out how far these graphs can be trusted, since
the first `extra' line in figure \ref{fig:5}a appears to descend from
level 6 (the first extra null state in $R_{(2,2)}$ for $t=4/5$)
first to level 4 
(the first extra null state in $R_{(3,1)}$ for $t=4/5$)
and further.
If this level really dropped below energy 4 then we
would have trouble identifying the spectrum as that of the boundary
condition $(3,1)$. 
One way to judge how far these graphs can be trusted is to see how the
pattern changes with the truncation level.
In figure \ref{fig:6} we plot the normalised energy gaps
$(E_i(r) - E_0(r))/(E_1(r)-E_0(r))$ for $\log|r|>0$ in the case 
$t=0.8$ for truncation 12, and on top of this we plot
the first `extra' line for $t=0.8002$ for truncation levels
8, 10, 12 and 14.
(It is important to note that the lines crossings in $M(4,5)$ are
absent for $M(4001,5000)$, but that the gaps between the lines are
so small that one can easily identify the `extra lines' that we plot here.)
It is clear that the spectrum is not really stable
for $\log|r|>2$, and that the crossing of the level 4 by the extra
line is never part of the stable spectrum -- hence one could easily
believe that the first extra line will really join the other lines at
energy level 4 after truncation effects are removed.

This discussion suggests that for $t>4/5$ the endpoint of the
$(22)\pm\ftt$ flows are unchanged from those of the unitary
models, and furthermore it appears that the spectrum stays
real for all $4/5\leq t < 1$, although we do not have any arguments to
support this. 
For $t<4/5$, we find that the spectrum develops large imaginary parts,
and this makes the identification of the IR fixed point much harder,
and so we shall not attempt to say anything more about this regime.

\resection{Conclusions}
\label{sec:conc}

We have shown how one may take the \clim\ limit of the $A$--type minimal
models with \clt\  with a boundary, and that in this limit the boundary
condition of type $(rr')$ splits into a superposition of $\min(r,r')$
fundamental boundary conditions.

This leads to a simple heuristic picture for the perturbations of the
\clt\  minimal models by the boundary fields $\f_{rr}$ of `low'
conformal weight. We have checked that this picture appears to be
correct in various unitary models through use of the TCSA method.
We have argued that the pattern of flows changes smoothly with central
charge (modulo the omission of lines corresponding to null states) and
that for the perturbation by $\ftt$
the IR endpoints of the nonunitary models 
with $t\geq 4/5$, that is $c\geq 7/10$, 
appear to be the same as those of the unitary models.

We have also found good evidence in the models
$M(4,5)$, $M(6,7)$ and $M(8,9)$
that the perturbation of the boundary condition $(2p,2p+1)$ by the
field $\f_{2p-1,2p-1}\equiv \f_{12}$
is integrable (as one would expect on general grounds \cite{GZ}).
This clearly deserves further investigation, as they may be amenable
to an exact analysis through non-linear integral techniques.

Questions which we also plan to consider in the future are whether
there are any new features in the $D$-type models, and whether we
can identify a bulk model for which the \ceq\ boundary conditions
$(\hat r)$ we have found are the natural boundary conditions.
This last point will be addressed in \cite{RW}.

Finally, in a recent paper \cite{ZZ}, Zamolodchikov and Zamolodchikov
considered Liouville theory with $c\geq 25$,
found that the boundary conditions are naturally 
labelled $(m,n)$, and noted that the subset $(1,n)$ has a special role.
It would be interesting to see if there is any relation to the \ceq\
boundary conditions presented here. 

\medskip

\noindent{\bf Acknowledgements}
{
\parindent 0pt
\parskip 3pt
\raggedright

The work was supported in part by EPSRC grant GR/K30667.
KG thanks the EPSRC for a research studentship.
We would like to thank 
D.~Bernard, P.E.~Dorey, A.~Recknagel, H.~Saleur, G.~Tak\'acs,
C.~Schweigert, R.~Tateo and J.-B.~Zuber for helpful discussions.
}

\bigskip
\bigskip
\bigskip
\medskip
\medskip

\appendix
\section*{Appendices }
\refstepcounter{section}
\label{app}

\subsection{A--series boundary structure constants}
\label{sec:bsc}

In \cite{Runk1} it was argued that the boundary structure constants of
A--series minimal models are given by the Fusion-matrix $\textsf{F}$, which
describes the transformation behaviour of conformal blocks. The result
is most simply put as
\be
  \tilde C^{(abc)}_{ij}{}^k = \F iacjbk 
\;.
\ee
However, this solution has to be rescaled to match the
normalisation in \eref{mm:norm}. To investigate the
$\eps\rightarrow 0$ limit of this expression it is convenient to have
an explicit expression for the structure constants. This is given in
the next section.

\subsubsection{Explicit minimal model F-matrix}
\label{sec:ef}

Consider the minimal model $M(p,p')$. Let $t{=}p/p'$ and
$d_{rs}{=}r{-}st$.
We want to find the matrix connecting the conformal blocks occurring
in the $x{\rightarrow}0$ and 
$x{\rightarrow}1$ expansion of the chiral correlator
$\cev{\phi_I}\phi_J(1)\phi_K(x)\vec{\phi_L}$. 
The indices are given by
Kac-labels 
  $I{=}(i,i')$, $J{=}(j,j')$, etc. 
  Let correspondingly $d_I{=}i{-}i' t$,  $d_J{=}j{-}j't$, etc.

\noindent From \cite{DFa85}(A.35) we find:
\bea
   b_{xy}(\alpha,\beta;\rho) 
&=& \displaystyle\prod_{g=1}^{y}
   \frac{ \Gamma(g\rho)\Gamma(\alpha{+}g\rho)\Gamma(\beta{+}g\rho) }{
          \Gamma(\rho)\Gamma(\alpha{+}\beta{-}2x{+}(y{+}g)\rho) }
\;,
\\
   m_{xy}(\alpha,\beta) 
&=& t^{2xy}
   \displaystyle\prod_{g=1}^{x}\prod_{h=1}^{y}
   \Big\{ (ht{-}g)(\alpha{+}ht{-}g)(\beta{+}ht{-}g)
   (\alpha{+}\beta{+}(y{+}h)t{-}(x{+}g)) \Big\}^{-1}
\;,
\\
   j(x,y;\alpha,\beta) 
&=& m_{xy}(\alpha,\beta) \cdot 
   b_{yx}(-\tfrac 1t\alpha,-\tfrac 1t\beta;\tfrac 1t) \cdot 
   b_{xy}(\alpha,\beta;t)
\;.
\eea
From \cite{FGP90}(3.5) we find:
\be
\ba{rl}
    a(s;x,y;\alpha,\beta,\gamma,\delta;\rho) 
&=  \displaystyle\sum_{h=\max(x,y)}^{\min(s,x+y-1)}
   \frac{\displaystyle\prod_{g=1}^{s-h} \sin\pi(\delta{+}(x{-}1{+}g)\rho)
        \displaystyle\prod_{g=1}^{h-y} \sin\pi(-\alpha{+}(s{-}x{+}g)\rho)}{
        \displaystyle\prod_{g=1}^{s-y}
          \sin\pi(-\alpha{+}\delta{+}(s{-}y{+}g)\rho)}
\\
&\times
   \frac{\displaystyle\prod_{g=1}^{y-1-(h-x)} \sin\pi(\beta{+}(s{-}x{+}g)\rho)
         \displaystyle\prod_{g=1}^{h-x} \sin\pi(\gamma{+}(x{-}1{+}g)\rho)}{
         \displaystyle\prod_{g=1}^{y-1} 
          \sin\pi(\beta{+}\gamma{+}(y{-}1{+}g)\rho)}
\\
&\times
   \displaystyle\prod_{g=1}^{h-x} 
   \frac{\sin\pi((x{+}y{-}h{-}1{+}g)\rho)}{\sin\pi(g\rho)}
   \displaystyle\prod_{g=1}^{s-h} 
   \frac{\sin\pi((h{-}y{+}g)\rho)}{\sin\pi(g\rho)}
\ea
\ee
Putting together \cite{DFa85}(4.1) and \cite{FGP90}(3.1) we find:
{
\renewcommand{\arraystretch}{1.7}
\be
\ba{rl}
&  {\renewcommand{\arraystretch}{1} \F IJKLPQ }
\\
=&  \ds
\frac{
  j\big(\frac 12({\ell} {-}i {-}1{+}q ), 
        \frac 12({\ell}'{-}i'{-}1{+}q') ; -d_I, d_L \big) 
     }{
  j\big(\frac 12(j {-}i {-}1{+}p ), 
        \frac 12(j'{-}i'{-}1{+}p') ; -d_I, d_J \big) 
} 
\;
\frac{
   j\big(\frac 12(j {+}k {-}1{-}q ), 
         \frac 12(j'{+}k'{-}1{-}q') ; d_J, d_K \big)
     }{
   j\big(\frac 12(k {+}{\ell} {-}1{-}p ), 
         \frac 12(k'{+}{\ell}'{-}1{-}p') ; d_K, d_L \big)} 
\\
\times&
  a\big(
    \frac 12(-i{+}j{+}k{+}{\ell}) ; 
    \frac 12( k{+}{\ell}{+}1{-}p), 
    \frac 12( j{+}k{+}1{-}q);
   -\frac 1t d_I,
   -\frac 1t d_J,
   -\frac 1t d_K,
   -\frac 1t d_L; 
    \frac 1t) 
\\
\times&
  a\big(
    \frac 12(-i'{+}j'{+}k'{+}{\ell}') ;
    \frac 12( k'{+}{\ell}'{+}1 {-}p'), 
    \frac 12( j'{+}k'{+}1 {-}q');
    d_I,d_J,d_K,d_L;t)
\ea
\ee
}

\subsubsection{The \clim\ limit of the boundary structure constants}
\label{app:bscs}

We shall take all boundary fields on a single b.c.\ $(\al)$ to be
canonically normalised,   
\be
  \f_i^{(\al\al)}(x) \, \f_j^{(\al\al)}(y)
= \frac{\delta_{ij}}{(x-y)^{2h_i}} 
 \;+\;
  {\textstyle \sum_k} \bc{\al}{\al}{\al}ijk\,
  (x-y)^{h_k-h_i-h_j} \,
  \f_k^{(\al\al)}(y)
 \;+\;
 \ldots
\;\;\;\;
  x>y
\;,
\label{eq:norm}
\ee
so that we only need give the structure constants
\be
  C^{(\al)}_{ijk}
\equiv
  \bc{\al}{\al}{\al}ijk
\;,
\label{eq:sccs}
\ee
which are cyclically symmetric (n.b. in the $A$--model with \clt\
these are also 
completely symmetric, but that is not the case for the D--type model).
However, when we consider boundary-condition changing operators, 
one cannot set both $\bc abaii1$ and $\bc babii1$ to one due to the
(normalisation independent) condition 
\cite{Lewe1,Runk1}
\be
  \bc abaii1 S_1^a 
= \bc babii1 S_1^b
\;,
\label{eq:ababab}
\ee
Instead we choose an ordering on the boundary conditions and set
\be
  \bc abaii1 
= \cases{
  1\;,           &~ $a<b$ \cr
  S_1^b/S_1^a\;, &~ $a>b$
  }
\ee
With these normalisations, one observes that the structure constants
have a well-defined limit, independent of the precise choice of
sequence or even if the minimal models in the sequence are unitary or
not. 

It turns out to be possible to find quite concise explicit
formulae for the limits of the boundary structure constants involving
only the boundary conditions $(\hat a)$ and the corresponding fields
$\hat\varphi_{r}^{(ab)} = \lim_\clim \f_{(1r)}^{(1a)(1b)}$, 
which we present in the next section.

\subsubsection{Limit of F for (1,p)-representations}
\label{sec:1p}

Define $\eps$ by $t=1-\eps$ and let $j,k,\ell,p,q\in\mathbb{N}$ fulfill
the conditions 
\be
\ba{cccccc}
 & |i{-}j|<p<i{+}j   \quad 
 & |k{-}\ell|<p<k{+}\ell \quad 
 & \hbox{where $i+j+p$ and $k+\ell+p$ are odd,}
\\
  &|i{-}l|<q<i{+}\ell \quad &|j{-}k|<q<j{+}k \quad &
  \hbox{where $i+\ell+q$ and $j+k+q$ are odd.}
  \label{1p:ranges}
\ea
\ee
We shall denote the limit of the F-matrix by
\be
   \Fh ijk\ell pq 
=  \lim_{\elim} \;
   \F{(1,i)}{(1,j)}{(1,k)}{(1,\ell)}{(1,p)}{(1,q)}
\;.
\label{1p:Flim}
\ee
Using the explicit expression for the minimal model F-matrix we find
that the limit \elim\ is finite if condition
\eref{1p:ranges} is fulfilled. 
Let 
$s=(-i{+}j{+}k{+}\ell)/2$,
$x=(k{+}\ell{+}1{-}p)/2$, 
$y=(j{+}k{+}1{-}q)/2$. 
Then 
\be
\ba{r@{}c@{}l}
    \Fh ijk\ell pq 
&=& 
    (-1)^{(s+k)(s+x+y+1)} 
    \frac{(k{+}\ell{-}x{-}1)!}{(k{+}\ell{-}2x)!} 
    \prod_{g=1}^{s-y}\frac{g!\,(i{+}g{-}2)!}
                           {(i{+}s{-}y{-}\ell{+}g)!\,(\ell{-}g)!} 
\\
&\times&
  \prod_{g=1}^{s-x}\frac{(i{+}s{-}x{-}j{+}g{-}1)!\,
  (j{-}g)!}{(g{-}1)!\,(i{+}g{-}2)!} 
  \prod_{g=1}^{x-1}\frac{(\ell{-}x{+}g)!\,(k{-}x{+}g)!}{(g{-}1)!\,
  (k{+}\ell{-}2x{+}g{+}1)!} 
  \prod_{g=1}^{y-1}\frac{g!\,(j{+}k{-}2y{+}g{-}1)!}{
  (j{-}y{+}g)!\,(k{-}y{+}g)!} 
\\[2mm]
&\times&
  \sum_{h=\max(x,y)}^{\min(s,x{+}y{-}1)}
  \frac{\prod_{g=1}^{s-h}(x{-}\ell{-}1{+}g)\prod_{g=1}^{x+y-1-h}(x{+}j{-}s{-}g)
  \prod_{g=1}^{h-x}(k{-}x{+}1{-}g)\prod_{g=1}^{h-y}(i{+}s{-}x{+}g)}{
  (h{-}x)!\,(h{-}y)!\,(x{+}y{-}h{-}1)!\,(s{-}h)!}
\;.
\ea\ee
For the indices in the range \eref{1p:ranges} the arguments of the
factorials are always non-negative. To normalise the structure
constants we need the F-matrix elements corresponding to the two-point
functions of boundary fields. Let $n=(a{-}b{+}i{+}1)/2$, then 
\be
   \Fh iaaib1 
=  \frac{b}{a{-}n{+}i} \; 
   \frac{(a{-}n)!\,(i{-}n)!\,(i{+}a{-}n{-}1)!}
        {(n{-}1)!\,(i{-}1)!\,(a{-}1)!\,(b{-}1)!} \;
   \bigg\{ \prod_{g=1}^{n-1} 
           \frac{(g{+}a{-}n)!\,(g{+}i{-}n)!}
                {  (g{-}1)!\,(g{-}1{+}b)!} 
   \bigg\}^2
\;.
\label{1p:Fnorm}
\ee
One can verify that \eref{1p:Fnorm} is positive provided the indices
are in their allowed ranges \eref{1p:ranges}. 
Since each sequence of $\widehat{\textsf{F}}$'s has a
well-defined limit \eref{1p:Flim}, taking the limit commutes with
addition and multiplication. It follows that the
$\widehat{\textsf{F}}$'s fulfill the pentagon identity. Define the
constants 
\be
  a\le b \;:\;\; A^{ab}_i = \bigg(\Fh iaaib1\bigg)^{1/2}>0 
\quad\hbox{ and }\quad 
  a>b \;:\;\; A^{ab}_i = \bigg(\Fh ibbia1\bigg)^{1/2}>0 
\;.
\ee
Then the structure constants in the $c=1$ theory are given by
\be
   \bch abcijk
 = \frac{A^{ac}_k}{A^{ab}_i\,A^{bc}_j} \Fh iacjbk
\;.
\label{1p:limbsc}
\ee
The normalisation has been chosen such that all structure constants are
real and $\bch aaaii1\!{=}1$. For boundary changing operators we have
$\bch abaii1\!{=}1$ if $a<b$ and $\bch abaii1\!{=}b/a$ if $a>b$. Since
the $\widehat{\textsf{F}}$'s fulfill the pentagon identity, the
structure constants \eref{1p:limbsc} solve the boundary sewing
constraint given in \cite{Lewe1}.

\newpage
\subsection{ The \clim\ limit of the (2,2)-boundary}
\label{app:ope}

In the limit \clim, there are two primary boundary fields of weight
zero,
\be
  1 =  \f_{11}
\;,\;\;
 \ph = \ftt
\;,
\ee
and three fields of weight one
\be
  \ps = \f_{13}
\;,\;\;
  \pb = \f_{31}
\;,\;\;
   d_3 
= \textstyle
  \lim\limits_\elim \frac{\sqrt{1{-}\eps}}{2\eps}\,
  {\f_{33}'}
\;,
\ee
where $'$ denotes the derivative along the boundary.
To work out the \OPES\ of these fields we need the structure constants
for \clt\ to order $O(\eps)$.
Since we normalise the fields on the $(2,2)$ boundary, we only need
give the cyclically symmetric structure constants \eref{eq:sccs}:
        %
        %
        %

{
\renewcommand{\arraystretch}{1.6}
\def\B{{(2,2)}}
\def\Y{{(3,3)}}
\def\U{{(1,3)}}
\def\V{{(3,1)}}
\def\id{{(1,1)}}

\be
\ba{r@{\;=\;}rcrcl}

  \bcs \B \U\U\U 
& -{\frac{2\sqrt 2}{\sqrt 3}} &+& \sqrt 6\,\eps &+& O(\eps^2)
\\

  \bcs \B \V\V\V 
& {\frac{2 \sqrt 2}{\sqrt 3}} &+& \sqrt 6\,\eps &+& O(\eps^2)
\\

    \bcs \B \Y\Y\Y 
&   \frac{2}{\sqrt 3} && &+& O(\eps^2)
\\

  \bcs \B \Y\Y\U 
&  && \frac{2\sqrt 2}{\sqrt 3}\,\eps &+& O(\eps^2)
\\

  \bcs \B \Y\Y\V 
& && \frac{2\sqrt 2}{\sqrt 3}\,\eps &+& O(\eps^2)
\\

  \bcs \B \Y\V\U 
& \frac{1}{\sqrt 3} && &+& O(\eps^2)
\\

\ea
\label{eq:22sc}
\ee
}%
\noindent%
We also need the \OPE\ of two generic primary boundary fields 
$\f_i$ of weight $h_i$ to a third:
{\renewcommand{\arraystretch}{2}
\[
\ba{c@{\;}c@{\;}c@{\;}l}

  \f_1(x) 
& \f_2(y) 
&=&
   (x-y)^{\Delta_1}\,
  C\,
  \f_3(y)
\;+\;
  (x-y)^{\Delta_1+1}\,
  \ds \frac{\Delta_2}{2h_3} 
  C\,
  \f'_3(y)
\;+\; \ldots
\;,
\;\;x>y

\ea
\]
\be
\hbox{where }
\Delta_1 \equiv h_3 - h_2 - h_1
\;,\;\;
\Delta_2 \equiv h_3 + h_1 - h_2
\;,\;\;
C \equiv C_{12}{}^3
\;.
\label{eq:ope}
\ee%
}%
Substituting the structure constants for the $(2,2)$ boundary
condition from equation \eref{eq:22sc} 
into equation \eref{eq:ope} and its derivatives,
and taking the limit \elim, we
obtain the \OPES\ in the \ceq\ model (all for $x>y$).
As an example, to obtain the \OPE\ of $d_3$ with itself, we consider
first that of $\ph$ with itself, for \clt:
\be
\ba{r@{\;\,}c@{\;\,}l}
  \ph(x)\;\ph(y)
&=&
  (x{-}y)^{-4\eps^2}
\;+\;
  \frac{2\sqrt 2}{\sqrt 3}\eps
  (x{-}y)^{1-2\eps}\,\ps(y)  
\;+\;
  \frac{2\sqrt 2}{\sqrt 3}\eps
  (x{-}y)^{1+2\eps}\,\pb(y)
\\[3mm]
&&
\;\;\;\;\;+\;
  \frac{2}{\sqrt 3}
  (x{-}y)^{-2\eps^2}\,\ph(y)
\;+\;
  \frac{1}{\sqrt 3}
  (x{-}y)^{1-2\eps^2}\,\ph'(y)
\;+\ldots
\;,\;\; x>y\;,
\ea
\label{eq:phph}
\ee
where we have dropped less singular terms and  terms of order $O(\eps^3)$.
Taking the limit \elim\ of this equation, we recover the first eqn.\ of
\eref{eq:ph-opes}, and taking the limit of the $x$ and $y$ derivatives,
we recover 
the last eqn.\ of \eref{eq:wt1}, e.g.
\be
\ba{r@{\;}c@{\;}l}
  d_3(x)\; d_3(y)
&=&
  \lim\limits_\elim \;
  \frac{1}{4\eps^2} \;
  \ph'(x)\; \ph'(y)
\\
&=&
  \lim\limits_\elim \;
  \frac{1}{4\eps^2}\Big(
    -4\eps^2(x{-}y)^{-2}
\;+\;
  \frac{2\sqrt 2}{\sqrt 3}\eps
  (2\eps)(x{-}y)^{-1}\,\ps(y)  
\\[3mm]
&&
\;\;\;\;\;\;\;+\;
  \frac{2\sqrt 2}{\sqrt 3}\eps
  (-2\eps)(x{-}y)^{-1}\,\pb(y)
\;+\;
  \frac{2}{\sqrt 3}
  (-2\eps^2)(x{-}y)^{-2}\,\ph(y)
\;+\; \ldots
  \Big)
\\[3mm]
&=&
    -\; \frac{1}{(x{-}y)^2}(1 + {\frac{1}{\sqrt3}}\ph(y))
  \;+\; \sqrt{\frac{2}{3}}\frac{1}{(x{-}y)}(\ps(y) - \pb(y))
  \;+\; O(1)
\;,
\;\; x>y
\;.
\ea
\ee
The \OPES\ involving the field $\ph$ are
both regular and exact: 
{\renewcommand{\arraystretch}{1.9}
\be
\ba{c@{\;}c@{\;}c@{\;}c@{\;}c@{\;}c@{\;}c@{\;}c}

  \ph(x) 
& \ph(y) 
&=&
   1 
&+& 
   \;\frac{2}{\sqrt 3}\,\ph(y)
\;,
\\

  \ph(x) 
& \ps(y) 
&=&
  \frac{1}{\sqrt 3}\,\pb(y)
&-& 
  \sqrt{\frac{2}{3}}\,d_3(y)
\;,
\\

  \ph(x) 
& \pb(y) 
&=&
  \frac{1}{\sqrt 3}\,\ps(y)
&-& 
  \sqrt{\frac{2}{3}}\,d_3(y)
\;,
\\

  \ps(x)
& \ph(y) 
&=&
  \frac{1}{\sqrt 3}\,\pb(x)
&+& 
  \sqrt{\frac{2}{3}}\,d_3(x)
\;,
\\

  \pb(x)
& \ph(y) 
&=&
  \frac{1}{\sqrt 3}\,\ps(x)
&+& 
  \sqrt{\frac{2}{3}}\,d_3(x)
\;,
\\

  \ph(x)
& d_3(y) 
&=&
  - \sqrt{\frac{2}{3}}\,\ps(y)
&-& 
    \sqrt{\frac{2}{3}}\,\pb(y)\;~
&\!\!\!\!+&
    {\frac{1}{\sqrt 3}}\,d_3(y)
\;,
\\

  d_3(x)
& \ph(y) 
&=&
    \sqrt{\frac{2}{3}}\,\ps(x)
&+& 
    \sqrt{\frac{2}{3}}\,\pb(x)\;~
&\!\!\!\!+&
    {\frac{1}{\sqrt 3}}\,d_3(x)
\;.

\ea
\label{eq:ph-opes}
\ee
}%
Note that the structure constants of these fields are no longer
cyclically symmetric -- for example 
$C_{d_3\,\phi\,\psi} = - C_{\phi\, d_3\,\psi} = \sqrt{2/3}$.
The \OPES\ of the weight one fields 
are more complicated; again we give them for $x>y$:
{\renewcommand{\arraystretch}{1.9}
\be
\ba{r@{\;}c@{\;}c@{\;}c@{\;}c@{\;}c@{\;}c@{\;\;}l}

  \ps(x) 
& \ps(y) 
&=&
   \frac{1}{(x{-}y)^2}
&-& 
   \;\frac{2\sqrt 2}{\sqrt 3}\,
   \frac{1}{(x{-}y)}\,
   \ps(y)
&+&O(1)\;,
\\

  \pb(x) 
& \pb(y) 
&=&
   \frac{1}{(x{-}y)^2}
&+& 
   \;\frac{2\sqrt 2}{\sqrt 3}\,
   \frac{1}{(x{-}y)}\,
   \pb(y)
&+&O(1)\;,
\\

  \ps(x) 
& \pb(y) 
&=&
   \frac{1}{\sqrt 3}\,
   \frac{1}{(x{-}y)^2}\,
   \f(y)
&-& 
   \;\frac{2}{\sqrt 3}\,
   \frac{1}{(x{-}y)}\,
   d_3(y)
&+&O(1)\;,
\\

  \pb(x) 
& \ps(y) 
&=&
   \frac{1}{\sqrt 3}\,
   \frac{1}{(x{-}y)^2}\,
   \ph(y)
&+& 
   \;\frac{2}{\sqrt 3}\,
   \frac{1}{(x{-}y)}\,
   d_3(y)
&+&O(1)\;,
\\

  \ps(x) 
& d_3(y) 
&=&
   ~\sqrt{\frac{2}{3}}\,
   \frac{1}{(x{-}y)^2}\,
   \ph(y)
&-& 
   \frac{2}{\sqrt 3}\,
   \frac{1}{(x{-}y)}\,
   \pb(y)
&-&
   \sqrt{\frac{2}{3}}\,
   \frac{1}{(x{-}y)}\,
   d_3(y)
\;+O(1)\;,
\\

  \pb(x) 
& d_3(y) 
&=&
   ~\sqrt{\frac{2}{3}}\,
   \frac{1}{(x{-}y)^2}\,
   \ph(y)
&+& 
   \frac{2}{\sqrt 3}\,
   \frac{1}{(x{-}y)}\,
   \ps(y)
&+&
   \sqrt{\frac{2}{3}}\,
   \frac{1}{(x{-}y)}\,
   d_3(y)
\;+O(1)\;,
\\

  d_3(x) 
& \ps(y) 
&=&
   -\sqrt{\frac{2}{3}}\,
   \frac{1}{(x{-}y)^2}\,
   \ph(y)
&+& 
   \frac{2}{\sqrt 3}\,
   \frac{1}{(x{-}y)}\,
   \pb(y)
&-&
   \sqrt{\frac{2}{3}}\,
   \frac{1}{(x{-}y)}\,
   d_3(y)
\;+O(1)\;,
\\

  d_3(x) 
& \pb(y) 
&=&
   -\sqrt{\frac{2}{3}}\,
   \frac{1}{(x{-}y)^2}\,
   \ph(y)
&-& 
   \frac{2}{\sqrt 3}\,
   \frac{1}{(x{-}y)}\,
   \ps(y)
&+&
   \sqrt{\frac{2}{3}}\,
   \frac{1}{(x{-}y)}\,
   d_3(y)
\;+O(1)\;,
\\

  d_3(x) 
& d_3(y) 
&=&
\multicolumn{5}{l}{\hbox to 0pt{\hbox{$
   \!\!
-  \;\frac{1}{(x{-}y)^2}
   (\,
   1+ \frac{1}{\sqrt 3}\ph(y)
   \,)
\;+\;
   \sqrt{\frac{2}{3}}\,
   \frac{1}{(x{-}y)}\,
   ( \ps(y) - \pb(y) )
\;+\;O(1)\;.$}}}

\ea
\label{eq:wt1}
\ee
}

\subsection{The \clim\ limit of the $(2,p)$ boundary condition}
\label{app:2p}

There are two fields of weight zero on the \clim\ limit of the $(2,p)$
b.c., namely $\f_{11}$ and $\f_{33}$.
From section \ref{app:bscs} we have
\be
  \bc {(2p)}{(2p)}{(2p)} {(33)}{(33)}{(33)} 
= \frac{2}{\sqrt{p^2-1}}
\;,
\ee
which gives the \OPE
\be
  \f_{33}(x) \; 
  \f_{33}(y) 
= \f_{11}(y) 
  \;+\;
  \tfrac{2}{\sqrt{p^2-1}}  \f_{33}(y) 
\;.
\ee
From this we deduce that the projectors are
\bea
    \hP_a 
&=& \tfrac{p-1}{2p}\f_{11} 
    \;+\; 
    \tfrac{\sqrt{p^2-1}}{2p}\f_{33} 
\;,
\\
    \hP_b 
&=& \tfrac{p+1}{2p}\f_{11} 
    \;-\;
    \tfrac{\sqrt{p^2-1}}{2p}\f_{33} 
\;.
\eea
These can be inverted to give
\bea
    \f_{11} 
&=& \hP_a  
    \;+\; 
    \hP_b
\;,
\\
    \f_{33} 
&=& \sqrt{\tfrac{p+1}{p-1}} \hP_a
    \;-\;
    \sqrt{\tfrac{p-1}{p+1}} \hP_b
\;.
\eea
In the case of the (22) boundary, 
$\hP_a$ projects onto the $(\hat 1)$
boundary and $\hP_b$ on the $(\hat 3)$ boundary,
and in general
$\hP_a$ projects onto the $(\hat {p-1})$
boundary and $\hP_b$ on the $(\hat {p+1})$ boundary,

\subsection{The \clim\ limit of the $(3,p)$ boundary condition}
\label{app:3p}

There are three fields of weight zero on the \clim\ limit of the
$(3,p)$ b.c., namely $\f_{11}$, $\f_{33}$ and $\f_{55}$.
From section \ref{app:bscs} we have
\be
\ba{rclrcl}
    A
\equiv 
    \bc {(3p)}{(3p)}{(3p)}{(33)}{(33)}{(33)} 
&=& \sqrt{\frac{3}{2}}\frac{1}{{\sqrt{p^2-1}}} 
\;,\;\;
&
    B
\equiv
    \bc {(3p)}{(3p)}{(3p)}{(33)}{(33)}{(55)}  
&=& \frac{1}{\sqrt{2}}\sqrt{\frac{p^2-4}{p^2-1}} 
\;,
\\[4mm]
    C
\equiv
    \bc {(3p)}{(3p)}{(3p)}{(55)}{(55)}{(33)} 
&=& 3\sqrt{\frac{3}{2}}\frac{1}{{\sqrt{p^2-1}}} 
\;,\;\;
&
    D
\equiv
    \bc {(3p)}{(3p)}{(3p)}{(55)}{(55)}{(55)} 
&=&  -\frac{1}{\sqrt{2}}\frac{p^2-16}{\sqrt{(p^2-4)(p^2-1)}} 
\;.
\ea
\ee
which gives the opes
\bea
    \f_{33}\; \f_{33} 
&=& \f_{11} \;+\; A \f_{33} \;+\; B \f_{55} 
\;,
\\
    \f_{33}\; \f_{55} 
&=& B \f_{33} \;+\; C \f_{55} 
\;,
\\
    \f_{55}\; \f_{55} 
&=& \f_{11} \;+\; C \f_{33} \;+\; D  \f_{55} 
\;.
\eea
From these we deduce that the projectors are
\bea
    \hP_a  
&=& \tfrac{p-2}{3p} \f_{11} 
\;+\; \tfrac{p-2}{p}\sqrt{\tfrac{p+1}{6(p-1)}} \f_{33} 
\;+\; \tfrac{1}{3p}\sqrt{\tfrac{(p+1)(p^2-4)}{2(p-1)}} \f_{55}  
\;,
\\
    \hP_b  
&=& \tfrac{1}{3} \f_{11} 
\;+\; \sqrt{\tfrac{2}{3(p^2-1)}}  \f_{33} 
\;-\; \tfrac{1}{3}\sqrt{\tfrac{2(p^2-4)}{(p^2-1)}} \f_{55}  
\;,
\\
    \hP_c 
&=& \tfrac{p+2}{3p} \f_{11} 
\;-\; \tfrac{p+2}{p}\sqrt{\tfrac{p-1}{6(p+1)}} \f_{33} 
\;+\; \tfrac{1}{3p}\sqrt{\tfrac{(p-1)(p^2-4)}{2(p+1)}} \f_{55}  
\;.
\eea
These can be inverted to give
\bea
    \f_{11} 
&=& \hP_{a} \;+\; \hP_b \;+\; \hP_{c}
\;,
\\
    \f_{33} 
&=& 
      \sqrt{\tfrac{3(p+1)}{2(p-1)}} \hP_{a}
\;+\; \sqrt{\tfrac{6}{p^2-1}}       \hP_{b} 
\;-\; \sqrt{\tfrac{3(p-1)}{2(p+1)}} \hP_c 
\;,
\\ 
    \f_{55} 
&=& 
      \sqrt{\tfrac{(p+2)(p+1)}{2(p-2)(p-1)}} \hP_{a} 
\;-\; \sqrt{\tfrac{2(p^2-4)}{p^2-1}}        \hP_b
\;+\; \sqrt{\tfrac{(p-2)(p-1)}{2(p+2)(p+1)}} \hP_c 
\;.
\eea
In the case of the (33) boundary, 
$\hP_a$, $\hP_b$ and $\hP_c$ project onto the 
$(\hat 1)$, $(\hat 3)$ and $(\hat 5)$ boundaries respectively.

\medskip
\medskip
\small
\renewcommand\baselinestretch{0.95}


\begin{thebibliography}{99}
\raggedright
\parskip 2pt

\bibitem{Affleck}
I.~Affleck,
{\em Edge Critical Behaviour of the 2-Dimensional Tri-critical Ising Model,}
\newline
{\em J.~Phys.}~{\bf A33} (2000) 6473-6480, 
{\tt cond-mat/0005286.}

\bibitem{BPTWa1}
Z. Bajnok, L. Palla, G. Tak\'acs and F. W\'agner,
{\em
The $k$ folded sine-Gordon model in finite volume},
Nucl. Phys. {\bf B587} (2000) 585-618, {\tt hep-th/0004181}.

\bibitem{BPTWa2}
Z. Bajnok, L. Palla, G. Tak\'acs and F. W\'agner,
{\em
Nonperturbative study of the two frequency sine-Gordon model},
{\tt hep-th/0008066}.

\bibitem{BPPZ1}
R.E. Behrend, P.A. Pearce, V.B. Petkova and J.-B. Zuber,
{\it 
Boundary Conditions in Rational Conformal Field Theories,}
Nucl. Phys. {\bf B570} (2000) 525--589,
\newblock {\tt hep-th/9908036.}

\bibitem{BPPZ2}
R.E. Behrend, P.A. Pearce, V.B. Petkova and J.-B. Zuber,
{\it 
On the classification of bulk and boundary conformal field theories,}
Phys.\ Lett.\ {\bf B444} (1998) 163--166,
\newblock {\tt hep-th/9809097.}


\bibitem{BPPZ3}
R.E. Behrend, P.A. Pearce, V.B. Petkova and J.-B. Zuber,
{\it 
Integrable boundaries, conformal boundary conditions and A-D-E fusion
rules,}
\newblock
J.~Phys.~{\bf A31} (1998) L763--L770,
\newblock {\tt hep-th/9807142.}

\bibitem{Cardy}
J.L.\ Cardy,
{\it Critical percolation in finite geometries},
\newline
J.~Phys.~{\bf A25} (1992) L201-L206,
{\tt hep-th/9111026}.

\bibitem{Card4}
J.L.\ Cardy,
{\it Boundary conditions, fusion rules and the Verlinde formula},
\newline
Nucl.~Phys.~{\bf B324} (1989) 581--596.

\bibitem{CLew1}
J.L. Cardy and D.C.\ Lewellen,
{\it Bulk and boundary operators in conformal field theory},
\newline
Phys.\ Lett.\ {\bf B259} (1991) 274--278.

\bibitem{Ybk}
Ph. Di Francesco, P. Mathieu, and D. S\'en\'echal,
{\em Conformal Field Theory},
Springer {\em Graduate texts in contemporary physics} 1997.


\bibitem{DPTWa1}
P.\ Dorey, A.\ Pocklington, R.\ Tateo and G.\ Watts,
{\it
TBA and TCSA with boundaries and excited states},
\newblock Nucl.\ Phys.\ {\bf B525} (1998) 641--663, {\tt hep-th/9712197}

\bibitem{DRTWa1}
P.~Dorey, I.~Runkel, R.~Tateo and G.~Watts,
{\em
$g$--function flow in perturbed boundary conformal field theories},
\newblock 
\newblock Nucl.\ Phys.\ {\bf B578} (2000) 85--122, {\tt hep-th/9909216}


\bibitem{DFa85}
 Vl.S.~Dotsenko, V.A.~Fateev,
{\em Four-point correlation functions and the operator algebra
     in 2d conformal invariant theories with central charge $c{\le}1$},
Nucl.~Phys.~{\bf B251} (1985) 691--734.

\bibitem{CS}
J.~Fuchs and C.~Schweigert,
{\em Solitonic sectors, conformal boundary conditions and
three-dimensional topological field theory},
talk given by CS at "Nonperturbative Quantum Effects 2000", 
PRHEP-tmr2000/039, {\tt hep-th/0009111}.

\bibitem{FGP90}
 P.~Furlan, A.Ch.~Ganchev, V.B.~Petkova,
{\em Fusion matrices and $c{<}1$ (quasi) local conformal
     field theories},
 Int.~J.~Mod.~Phys.~{\bf A5} (1990) 2721--2735.

\bibitem{GZ} 
S.~Ghoshal, A.B.~Zamolodchikov,
{\em Boundary S matrix and boundary state in two-dimensional integrable
quantum field theory},
Int.~J.~Mod.~Phys.~{\bf A9} (1994) 3841-3886, 
erratum 
Int.~J.~Mod.~Phys.~{\bf A9} (1994) 4353,
{\tt hep-th/9306002}

\bibitem{GRW1} 
K.~Graham, I.~Runkel and G.M.T.~Watts, 
{\em
Renormalisation group flows of boundary theories},
talk given by GMTW at "Nonperturbative Quantum Effects 2000", 
PRHEP-tmr2000/040,
{\tt hep-th/0010082}.

\bibitem{next} 
K.~Graham and G.M.T.~Watts, {\em in preparation.}

\bibitem{StAubin}
E.~Lapalme and Y.~Saint-Aubin,
{\it 
Crossing probabilities on same spin clusters in the two-dimensional Ising
model},
\newblock
{\tt hep-th/0005104}.

\bibitem{Lewe1}
D.C.\ Lewellen,
{\it Sewing constraints for conformal field theories on surfaces with
boundaries},
\newblock Nucl.\ Phys.\ {\bf B372} (1992) 654--682.

\bibitem{PZube}
V.B.~Petkova and J.-B.~Zuber,
\newline
{\it The many faces of Ocneanu cells},
{\tt hep-th/0101151};
\newline
{\it BCFT: from the boundary to the bulk},
talk given by JBZ at "Nonperturbative Quantum Effects 2000", 
PRHEP-tmr2000/038,
{\tt hep-th/0009219}.

\bibitem{Pol}
J.~Polchinski,
{\it TASI Lectures on D-Branes},
{\tt hep-th/9611050}.

\bibitem{RRSch1} 
A. Recknagel, D. Roggenkamp and V. Schomerus,
\newline
{\em On relevant boundary perturbations of unitary minimal models},
\newline
Nucl.~Phys.~{\bf B588} (2000) 552-564,
{\tt hep-th/0003110}.

\bibitem{Runk1}
I. Runkel,
\newblock
{\it 
Boundary structure constants for the A-series Virasoro minimal
models},
\newblock Nucl.~Phys.~{\bf B549} (1999) 563-578, {\tt hep-th/9811178}.

\bibitem{RW}
I.~Runkel and G.M.T.~Watts, in preparation.

\bibitem{Watts}
G.M.T Watts,
{\it 
A crossing probability for critical percolation in two-dimensions},
\newline
J.~Phys.~{\bf A29} No 14 (1996) L363, 
{\tt cond-mat/9603167}. 

\bibitem{ZZ}
A.B.~Zamolodchikov and Al.B.~Zamolodchikov,
{\it
Liouville field theory on a pseudosphere},
{\tt hep-th/0101152}.


\end{thebibliography}
\end{document}